# Antioxidant activity and toxicity study of cerium oxide nano-particles stabilized with innovative functional copolymers

**Geoffroy Goujon[1], Victor Baldim[2], Caroline Roques[3], Nicolas Bia[4], Johanne Seguin[3], Bruno Palmier[1], Alain Graillot[4], Cédric Loubat[4], Nathalie Mignet[3], Isabelle Margaill[1*], Jean-François Berret[*2] and Virginie Beray-Berthat[*5]**

[1]Université de Paris, Inserm UMR_S1140, "Innovative Therapies in Haemostasis", Paris, France;
[2]Université de Paris, CNRS UMR 7057, Matière et systèmes complexes, 75013 Paris, France
[3]Université de Paris, UTCBS (Unité de Technologies Chimiques et Biologiques pour la Santé), CNRS UMR8258, – Inserm U1267, – Inserm, 4 avenue de l'observatoire, F-75006, Paris, France
[4]SPECIFIC POLYMERS, ZAC Via Domitia, 150 Avenue des Cocardières, F-34160 Castries, France -
[5]Université de Paris, CNRS ERL 3649 "Pharmacologie et thérapies des addictions", Inserm UMR-S 1124 T3S "Environmental Toxicity, Therapeutic Targets Cellular Signaling an biomarkers", 45 rue des Saints Pères, F-75006 Paris, France

**Abstract:** Oxidative stress, which is one of the main harmful mechanisms of pathologies including ischemic stroke, contributes to both neurons and endothelial cell damages, leading to vascular lesions. Although many antioxidants have been tested in preclinical studies, no treatment is currently available for stroke patients. Since cerium oxide nanoparticles (CNPs) exhibit remarkable antioxidant capacities, our objective is to develop an innovative coating to enhance CNPs biocompatibility without disrupting their antioxidant capacities or enhance their toxicity. This study reports the synthesis and characterization of functional polymers and their impact on the enzyme-like catalytic activity of CNPs. To study the toxicity and the antioxidant properties of CNPs for stroke and particularly endothelial damages, *in vitro* studies are conducted on a cerebral endothelial cell line (bEnd.3). Despite their internalization in bEnd.3 cells, coated CNPs are devoid of cytotoxicity. Microscopy studies report an intracellular localization of CNPs, more precisely in endosomes. All CNPs reduces glutamate-induced intracellular production of ROS in endothelial cells but one CNP significantly reduces both the production of mitochondrial superoxide anion and DNA oxidation. *In vivo* studies report a lack of toxicity in mice. This study therefore describes and identifies biocompatible CNPs with interesting antioxidant properties for ischemic stroke and related pathologies.

## 1. Introduction

Oxidative stress results from an imbalance between the production of reactive oxygen/nitrogen species (ROS/RNS) and endogenous antioxidant systems. Oxidative stress is a major contributor to the pathogenesis of ischemic stroke (due to the occlusion of a brain vessel).[1–3] Besides neurons, post-ischemic oxidative stress also contributes to endothelial cell damages, leading to vascular lesions and subsequent cerebral hemorrhages, which are associated with a poor prognosis for stroke patients.[4] Although many antioxidants have been tested in preclinical studies and, for a few of them, in clinical trials, there is no antioxidant treatment currently available for stroke.[5–8] Cerium oxide nanoparticles (CNPs), also called nanoceria, hold wide-ranging antioxidant capacities. They indeed scavenge both ROS and RNS such as hydroxyl radical (HO•)[9] and nitric oxide (•NO)[10] while mimicking the antioxidant enzymes superoxide dismutase (SOD) and catalase (CAT).[11–15] Interestingly, the antioxidant activity of CNPs has been associated with a protective effect not only on neurons[16,17], but also on endothelial cells.[18] Beneficial effects of CNPs have been reported in preclinical models of various pathologies such as cancer,[19–22] cardiovascular injuries[23,24] or diseases of the central nervous system[16,25] such as Alzheimer's disease.[26] With regards to brain ischemia, neuroprotection by CNPs has been first shown *ex vivo* on hippocampal slices submitted to hypoxia.[27] In a rat model of focal cerebral ischemia, CNPs were shown to reduce ROS production and the lesion's volume.[28] More recently, CNPs either loaded with an antioxidant, edaravone, or combined with a targeting ligand efficiently decreased the outcomes of ischemic strokes in rats.[18,29,30]

Despite these encouraging results, cerium oxide nanoparticles suffer from certain limitations. Their clinical translation requires a significant improvement of their surface properties, particularly through the development of an appropriate functionalization. Control of interfacial properties is a central issue in nanoparticle's development to increase their biocompatibility, stability and stealthiness in biological media. In particular, uncoated metal oxide nanomaterials are unstable in such media[31], thus leading to a poor pharmacodynamic profile. In addition, nanoceria's coating strongly influences cellular uptake and cytotoxicity.[32,33] The objective of our study is to develop coated CNPs displaying increased stability while retaining their antioxidant properties and decreasing their cytotoxicity.

One of the most promising surface functionalization methods makes use of polymers and co-assembly techniques.[34,35] This method takes advantage of the library of polymer architectures and binding agents synthesized polymer and coordination chemistry. Many types of polymers, natural or synthetic have indeed been used on cerium oxide nanoparticles : citrate/ poly(acrylic) acid[32,33], dextran/ poly(acrylic) acid [36], oleic acid – PAAOA – PMAOA[37], PSPM – PMETAC[38], PVA - PBCA – PGLA[39,40], DNC – PNC – ANC[41], heparin[42]. Recently, our group proposed a coating strategy based on functional statistical graft copolymers where phosphonic acids (a strongly binding agent to metals)[43,44] and poly(ethylene glycol) (PEG) chains are covalently grafted to a poly(methyl methacrylate) backbone. Copolymers with multiple phosphonic acids provide resilient coatings and long-term stability (> months). In addition, they are well suited to a broad range of metal oxide particles, including cerium, aluminum, iron and titanium oxides.[45] Protein adsorption studies have highlighted that PEG densities around 0.5 $nm^{-2}$ and layer thickness about 10 nm (corresponding to PEG molecular weight of 2000 to 5000 g $mol^{-1}$) provide excellent serum protein resistance.[46] *In vivo* magnetic resonance imaging established the potential of phosphonic acid PEG polymer coatings to significantly prolong the pharmacokinetics of intravenously injected iron oxide nanoparticles in mice.[47] Last but not least, this approach paves the way to use amine terminated PEG terpolymers that can be later covalently functionalized with targeting or imaging moieties.

In this article, we describe the cerium oxide nanoparticles structural and antioxidant property, and the functional polymers used as coating. These polymers are statistical copolymers bearing on one hand phosphonic acids for binding to the cerium surface and on the other hand either PEGs or amine modified PEGs to reduce the mononuclear phagocyte system uptake and allow further functionalization[45,46,48]. The impact of the coating on the antioxidant properties of nanoceria was also examined *ex vitro*. To study the interest of these antioxidant properties in the context of stroke and specifically against vascular lesions, *in vitro* studies were conducted on cerebral endothelial cells (bEnd.3 cells). We assessed nanoceria toxicity, their cellular uptake and intracellular location in endothelial cells and their effect on ROS production. Finally, we investigated the *in vivo* toxicity and biodistribution of nanoceria in mice.

## 2. Results

### 2.1. Nanoceria coating and characterization

#### 2.1.1. Bare nanoceria particles

Cerium oxide nanoparticles were synthesized by thermo-hydrolysis of cerium nitrate salt under hydrothermal conditions at acidic pH.[49,50] The nanoceria structure was resolved using a combination of techniques including transmission electron microscopy (TEM), dynamic light scattering (DLS), X-ray photoelectron spectroscopy (XPS), wide-angle X-ray scattering (WAXS) and UV-Visible spectrometry. Figure 1a shows a TEM image of CNPs in the form of 2.9 nm crystallite agglomerates. The size distribution obtained leads to a median diameter of 7.8 nm and a dispersity of 0.17. Using static and dynamic light scattering, the hydrodynamic diameter $D_H$ and the molecular weight $M_w^{NP}$ of the particles were found at 9.8 nm and $3 \times 10^5$ g mol$^{-1}$.[45,51] Figure 1b displays XPS results obtained from bare CNP powder samples. The Ce3d XPS spectrum is decomposed into five pairs of peaks,[52,53] three being associated with the Ce$^{4+}$ ions and two with the Ce$^{3+}$ ions. The XPS data analysis on the peak assignment, on the determination of the binding energies and on the Ce$^{3+}$ fraction is provided in the **Supplementary Information S1** and leads to a Ce$^{3+}$-fraction of 14%. Note that with this Ce$^{3+}$ fraction, only one pair of Ce3d peaks (here the U'-V' indicated in red) is visible in the XPS intensity.

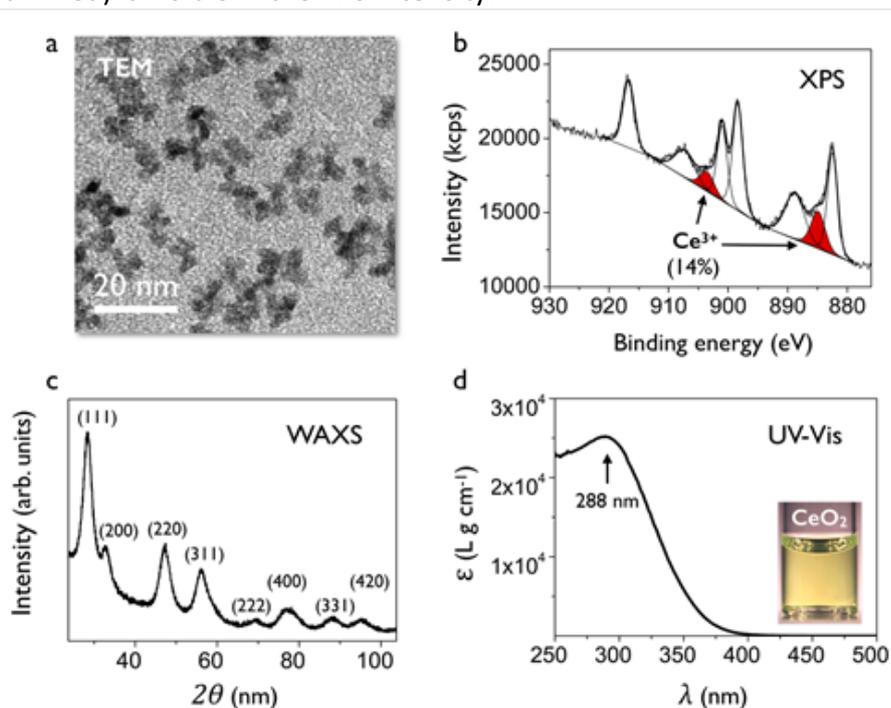

**Figure 1:** *Nanoceria characterization using a) transmission electron microscopy (TEM), b) X-ray photoelectron spectrometry (XPS), c) wide-angle X-ray scattering (WAXS) and d) CeO₂ dispersion absorptivity curve obtained from UV-Visible spectrometry. Inset: image of a concentrated CeO₂ dispersion. The nanoceria used in this work have the face-centered cubic fluorite-like structure and a Ce³⁺ fraction of 14%. The particles are small clusters (arrows in Figure 1a) composed of 2.9 nm crystallites. The cluster size has median diameter of 7.8 nm and a dispersity of 0.17 (the dispersity is defined as the ratio between the standard deviation and the average).*

The WAXS diffractogram in Figure 1c reveals that the nanoceria have a fluorite-like face-centered cubic structure. From the Rietveld MAUD analysis, the lattice constant was derived and found at 0.54151 nm. From the width of the Bragg reflections, the crystallite size was estimated at 2.9 nm. UV-Vis spectrometry performed on dispersions at different concentrations between 0.02 and 2 g L$^{-1}$ allowed to retrieve the absorptivity $\varepsilon(\lambda)$ as a function of the wavelength (Figure 1d). With decreasing $\lambda$, the $\varepsilon$-data reveal a maximum at 290 nm followed by a strong decrease down to 400 nm, in excellent agreement with earlier reports.[28,37,51]

### 2.1.2. Polymer synthesis and characterization

Three statistical copolymers were synthesized through free radical polymerization. The first one (P1) was obtained through the copolymerization of PEG methacrylate with a methacrylic monomer bearing a phosphonate group (MPh). Resulting copolymer is composed of an equimolar ratio (0.50:0.50) of phosphonic acid groups and methyl terminated PEGylated lateral chains. The PEGs have molecular weight 2000 g mol$^{-1}$, leading to the acronyms MPEG$_{2K}$-MPh where "M" refers to methacrylic nature of the co-monomers and "Ph" to phosphonic acid (Figure 2 and Table I). The synthesis was performed following a procedure previously described.[48,54] The two other polymers (P2 and P3) are terpolymers containing methyl-and amine-terminated PEG chains in addition to the phosphonic acid groups (Figure 2 and Table I). The amine terminated PEGs have molecular weight of 1000 or 2000 g mol$^{-1}$, leading to the acronyms MPEG$_{2K}$-MPEGa$_{1K}$-MPh and MPEG$_{2K}$-MPEGa$_{2K}$-MPh, where "a" refers to the amine terminal group. The proportions of PEGs, amine modified PEGs, and phosphonic acids are (0.35:0.15:0.50) for P2 and (0.07:0.43:0.50) for P3. With a PEGa$_{1K}$ as a co-monomer, it is expected that functional groups linked to the primary amine will be partially embedded in the PEG$_{2K}$ brush and be protected from non-specific binding. The polymers' molecular structures are illustrated in Figure 2. Details on the P2 and P3 terpolymer synthesis and their $^1$H NMR characterizations can be found in the materials and methods section and in **Supplementary Information S2**. The weight-averaged molecular weights $M_w^{Pol}$ were determined from static light scattering using the Zimm plots.[55] The P1, P2 and P3 copolymers are found with a molecular weight $M_w^{Pol}$ of 20300, 39500 and 29200 g mol$^{-1}$ respectively. Assuming a molar mass dispersity Đ = 1.8,[46,48] the number-averaged molecular weight $M_n^{Pol}$ is determined at 11300, 21900 and 16200 g mol$^{-1}$. From these values, the average number of phosphonic acids was estimated at 5.1, 7.7 and 6.6, confirming the presence of multiple functional groups. These later results are summarized in Table I.

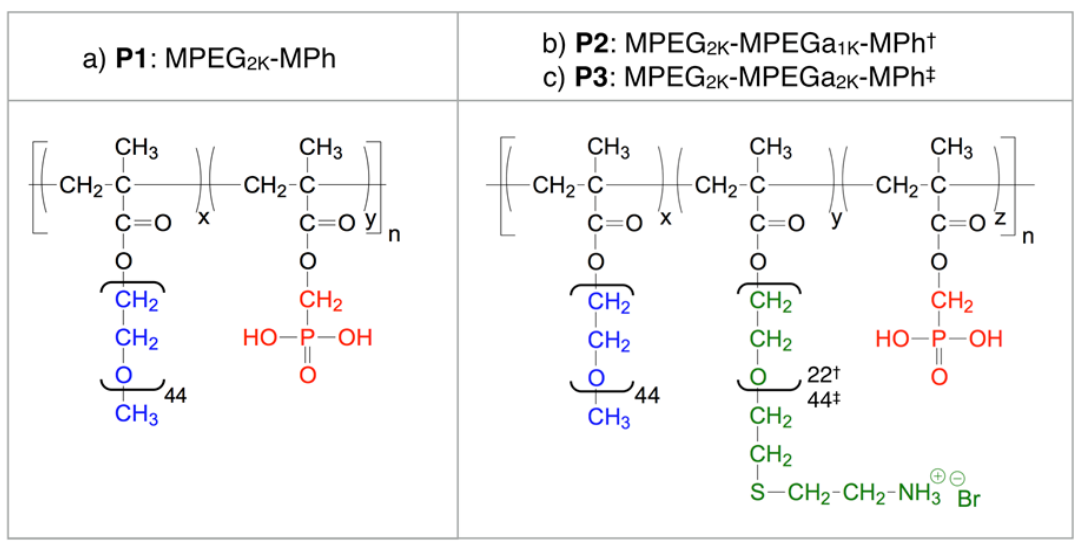

**Figure 2:** *Molecular structures of phosphonic acid-based copolymers and terpolymers synthesized in this work.* **a)** *The poly(poly(ethylene glycol) methacrylate-co-dimethyl(methacryooyloxy)methyl phosphonic acid) is a statistical copolymer where the repeating units have lateral methyl terminated $PEG_{2K}$ chains and lateral phosphonic acids in the proportion (0.50:0.50). It is abbreviated as P1: $MPEG_{2K}$-MPh.* **b)** *The statistical terpolymer P2: $MPEG_{2K}$-$MPEGa_{1K}$-MPh has for repeating units lateral methyl terminated $PEG_{2K}$ chains, lateral amine terminated $PEG_{1K}$ chains and lateral phosphonic acids in the molar proportions (0.35:0.15:0.50).* **c)** *As in b) with lateral amine terminated $PEG_{2K}$ chains and molar proportions (0.07:0.43:0.50). This polymer is abbreviated as P3: $MPEG_{2K}$-$MPEGa_{2K}$-MPh.*

**Table I:** *Molecular characteristics of the phosphonic acid PEGylated polymers and copolymers synthesized in this work.*

| Abbreviation | Polymer | $M_w^{Pol}$ (g mol$^{-1}$) | $M_n^{Pol}$ (g mol$^{-1}$) | Anchoring groups per polymer |
|---|---|---|---|---|
| P1 | $MPEG_{2K}$-MPh | 20300 | 11300 | 5.1 |
| P2 | $MPEG_{2K}$-$MPEGa_{1K}$-MPh | 39500 | 21900 | 7.7 |
| P3 | $MPEG_{2K}$-$MPEGa_{2K}$-MPh | 29200 | 16200 | 6.6 |

## 2.1.3. Nanoceria coating

The coating of nanoceria was carried out at acidic pH (1.5) to prevent aggregation. Moreover, it was found that below the critical mixing ratio of 1.5, well-dispersed coated particles were obtained whilst above, they associate and form large aggregates that eventually precipitate in solution[45]. In this study, the mixing ratio was defined as the mass ratio between particles and polymers in the mixed dispersion. These results are interpreted in the framework of the non-stoichiometric adsorption model we developed in the context of polymer coating.[51] This model assumes that the polymers adsorb spontaneously onto CNPs thanks to the phosphonic acid groups anchoring at the surface. The association is described as non-stoichiometric because the number of polymers adsorbed per particle depends on the mixing ratio. Dynamic light scattering was used to measure the polymer thickness $h$.[45,47] For the three polymer coats, we found $h$ = 9.1, 10.2 and 11.2 nm (Table II). These values are consistent with stretched PEG chains forming a polymer brush[56,57]. From the critical mixing ratio, the PEG density can be estimated, resulting in 0.3, 0.2 and 0.6 nm$^{-2}$ for P1, P2 and P3 coating respectively. Electrokinetic measurements using laser Doppler velocimetry and phase analysis light scattering mode show that the bare nanoceria are positively charged with a zeta potential $\zeta$ = + 21 mV, whereas coated particles are globally neutral (Table II). Concerning the colloidal stability, earlier studies have shown that nanoceria coated with phosphonic acid PEG copolymers provide resilient coatings and long-term stability, while avoiding protein adsorption (**Supplementary Information S3)**.

**Table II** – *Polymer coated cerium oxide nanoparticle hydrodynamic diameter $D_H$, polymer thickness $h$ and zeta potential $\zeta$ determined for particles dispersed in the acetate buffer 0.1M (pH 4.0).*

| Nanoparticles | $D_H$ (nm) | $h$ (nm) | $\zeta$ (mV) |
|---|---|---|---|
| CeO$_2$ bare | 9.0 | .. | +21 |
| CeO$_2$@P1 | 27.2 | 9.1 | +1.4 |
| CeO$_2$@P2 | 29.5 | 10.2 | -1.1 |
| CeO$_2$@P3 | 31.5 | 11.2 | +5.8 |

### 2.1.4. Antioxidant properties of nanoceria

The antioxidant capacities of nanoceria were evaluated *ex vitro* regarding SOD- and CAT-like activities, to assess the impact of multi-PEG coatings. The SOD-like activity $A_{SOD}$ is defined as the percentage of dismutated superoxide anions at the end of the assay. Figure 3a shows the $A_{SOD}$ for bare and coated nanoceria (P1:MPEG$_{2K}$-MPh, P2:MPEG$_{2K}$-MPEGa$_{1K}$-MPh and P3:MPEG$_{2K}$-MPEGa$_{2K}$-MPh) at 1, 10, 100 and 1000 µg ml$^{-1}$. Our results showed similar dismutation rates between bare and coated CNPs, which increased with concentration. The concentration results are interpreted in terms of a Langmuir-type adsorption isotherm where $A_{SOD}$ scales with the surface area concentration and with the Ce$^{3+}$ fraction.[13,15] The addition of a polymer coat on the particles did not alter their catalytic activity. The catalase-like activity of nanoceria was investigated at 1, 10, 100 and 1000 µg mL$^{-1}$. The CAT-like activity $A_{CAT}$ was defined as the percentage of decomposed H$_2$O$_2$ at the end of the reaction. $A_{CAT}$ increases sharply as a function of the nanoceria concentration and saturates at high dose (1000 µg mL$^{-1}$). When coated, a 20% reduction of the CAT-like activity of nanoceria is observed compared to bare CNPs, as shown for the data at 10 and 100 µg mL$^{-1}$. These results demonstrate that phosphonic acid PEG copolymers only slightly affect the CNP biomimetic catalytic activity, while increasing their long-term colloidal stability.

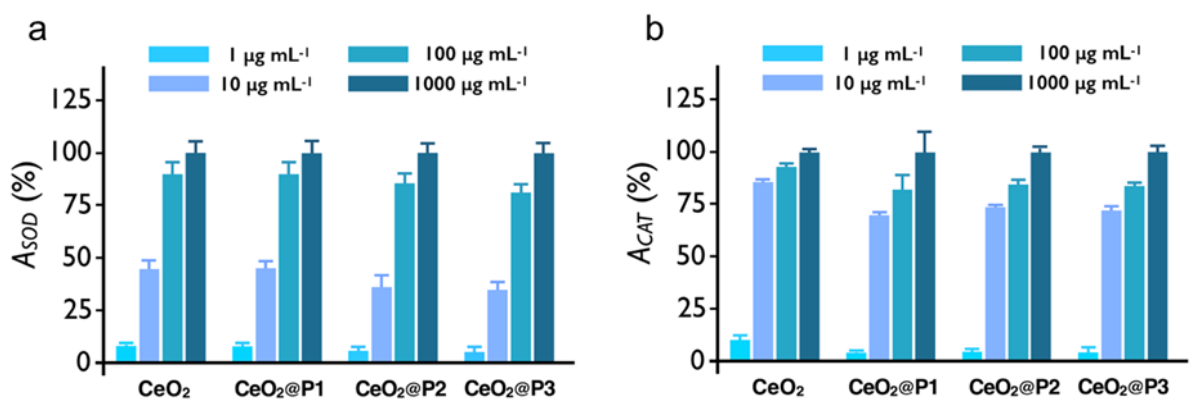

**Figure 3:** *Nanoceria superoxide dismutase- and catalase-like activities. The catalytic activity of bare nanoceria and nanoceria coated with P1 (MPEG$_{2K}$-MPh), P2 (MPEG$_{2K}$-MPEGa$_{1K}$-MPh) and P3 (MPEG$_{2K}$-MPEGa$_{2K}$-MPh) as a function of their concentrations was assessed against superoxide anions **(a)** and hydrogen peroxide **(b)**. Data are expressed as mean ± SEM.*

## 2.2. Effect of CNPs on murine cerebral endothelial cells bEnd.3

### 2.2.1. Toxicity of nanoceria

Metabolic activity and mortality of bEnd.3 cells were evaluated after 4-hour and 24-hour incubations with nanoceria (Figure 4). Glutamate, an excitatory neurotransmitter, whose involvement in cerebral post-ischemic oxidative stress and damage is well established, was used as a positive control for toxic effect: glutamate (100 mM) induced a loss in metabolic activity at both 4 hours and 24 hours (- 17% P < 0.01 and - 50% P < 0.001 respectively compared to control cells) and an increase in bEnd.3 cell mortality at 24 hours (+14% P < 0.01 *versus* control; Figure 4). With regards to CNPs, at lower concentrations (*i.e.* 10 and 100 µg mL$^{-1}$), neither the metabolic activity nor the mortality was modified compared with control cells, independently of the nanoparticle tested. At the highest concentration (*i.e.* 1000 µg mL$^{-1}$), CeO$_2$@P1 and CeO$_2$@P2 reduced the metabolic activity at 4 hours (-25% and -26% P < 0.01 respectively), and all CNPs induced a significant loss in metabolic activity at 24 hours (P < 0.01 and P < 0.001). However, the metabolic activity of

bEnd.3 cells incubated with 1000 µg mL$^{-1}$ of CeO$_2$@P1 or CeO$_2$@P3 during 24 hours was significantly higher compared to cells incubated with bare nanoceria at the same concentration, indicating that both P1 and P3 coatings are reducing the toxicity of CNPs. Only bare nanoparticles induced mortality as observed after both 4- and 24-hour incubation at the highest concentration (P < 0.001). At this concentration (*i.e.* 1000 µg mL$^{-1}$), the mortality observed with CeO$_2$@P1, CeO$_2$@P2 and CeO$_2$@P3 was significantly lower than bare CeO$_2$ (P < 0.001). The concentration of 1000 µg mL$^{-1}$ of CeO$_2$ is high compared to the concentrations used *in vitro* in the literature, *i.e.* most studies, notably on endothelial cells, used doses below 100 µg mL$^{-1}$.[18,28,58–60] With regards to the lower concentrations we used, our results are thus consistent with these studies which did not report any toxicity and/or decrease in viability on endothelial cells, primary or derived from cell lines, whether human or animal[58]. At the exception of Turco *et al.*[60] which reported a slight but significant reduction in cell viability at 100 µg mL$^{-1}$ of CNPs on HUVECs (human umbilical vein endothelial cells).

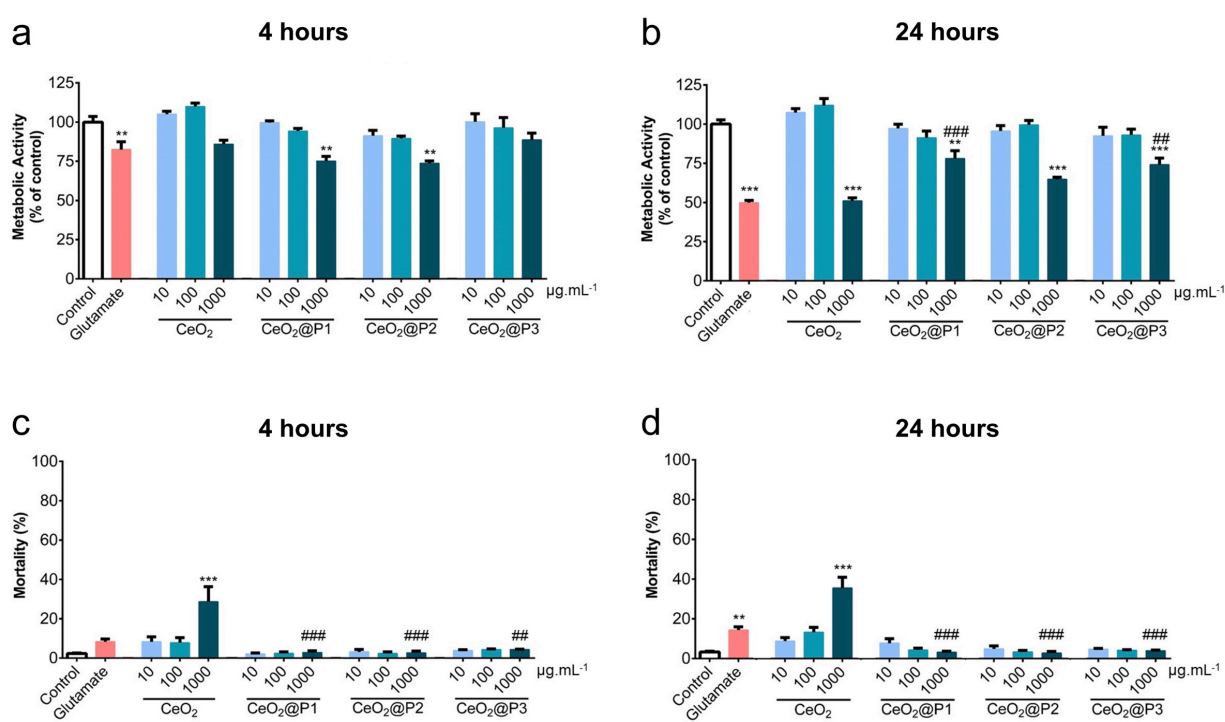

**Figure 4:** *Effect of nanoceria on metabolic activity and mortality of bEnd.3 cells. Toxicity of nanoceria was assessed measuring MTT metabolic activity (a, b) and trypan blue mortality (c, d), 4 hours (a, c) and 24 hours (b, d) after treatments. For the MTT assay (a, b), results are expressed in percentage of absorbance of control cells, and for the trypan blue assay (c, d), results are expressed in percentage of mortality. Glutamate was used at 100 mM. Data are expressed as mean ± SEM, n = 7-15 at 4 h and n = 5-13 at 24 h, \*\* P < 0.01, \*\*\* P < 0.001 versus control, ## P < 0.01, ### P < 0.001 versus bare nanoceria at 1000 µg mL$^{-1}$.*

### 2.2.2. Adsorption, internalization and intracellular location of nanoceria

Adsorption/internalization of nanoceria in bEnd.3 cells was first examined by quantification of cerium using ICP-OES (Inductively Coupled Plasma - Optical Emission Spectrometry) (Figure 5a). Our results showed that all CNPs associate with bEnd.3 cells independently of the concentration used, indicating that nanoceria are thus interacting with bEnd.3 cells after a 24-hour incubation, even at the lowest concentration (1 µg mL$^{-1}$). Whatever the coating, the detected amount of coated CNPs was drastically reduced compared to bare nanoceria (about 100 fold). This result highlights the stealthiness of coated CNPs compared to their bare counterpart. According to the literature and our data, coating strongly influences internalization[32,33]. However, Qiu *et al.*[38] reported only a slight decrease of cerium internalization with coated particles compared to our data. They only showed a 3-time reduction with a PSPM coating (3-sulfopropylmethacrylate with negative charges) and by 20% with PMETAC coating (2-(methacryloyloxy)ethyl-trimethylammoniumchloride) compared to bare CNPs after a 48-hour incubation.[38] These data underline the furtivity provided to the CNPs by our coating technique. Concerning flow cytometry analysis, our results are in line with data from the literature reporting uptake of nanoceria into endothelial cells within the first few hours of incubation.[18]

Finally, when the bEnd.3 cells were treated with glutamate, we did not observe any change in the internalization ratios (data not shown), which suggests that the influence of the coating is not modulated under these oxidative stress conditions. Additional analyses for adsorption/internalization were carried out through flow cytometry and confocal fluorescence microscopy. Considering the results of coated CNPs on cell toxicity, the concentration of 100 µg mL$^{-1}$ was selected for the following experiments. The primary amine groups were first quantified by spectrofluorometry using fluorescamine (**Supplementary Information S4**). Then, amino-PEG coated CNPs, *i.e.* CeO$_2$@P2 and CeO$_2$@P3, were labelled with cyanine-5 (Cy5). The amine free particles CeO$_2$@P1 were used as a control as no Cy5 is grafted on their surface. Flow cytometry analysis showed that cell basal auto-fluorescence was very low and was not modified by CeO$_2$@P1. Cyanine-fluorescence was detected in more than 95% of the cells incubated with CeO$_2$@P2-Cy5 and CeO$_2$@P3-Cy5 at 100 µg mL$^{-1}$ after a 4-hour incubation (Figure 5b). With regards to fluorescence intensity, a 4 to 6-fold increase was observed at 24 h compared to 4 h post treatment for CeO$_2$@P2-Cy5 and CeO$_2$@P3-Cy5 respectively ($P < 0.01$ and $P < 0.05$), indicating an increase in CNPs adsorption/internalization over time (Figure 5b).

Concerning the intracellular location, confocal fluorescence microscopy showed internalization of both cyanine-labelled CNPs CeO$_2$@P2 and CeO$_2$@P3 in bEnd.3 cells (Figure 5c). Observations at higher magnification indicated a perinuclear location. The localization of CNPs in different cell types has been studied by others using fluorophore-labelled CNPs. All studies showed localization only in the cytoplasm[18,38,61–63], including in endothelial cells[18,59], which is consistent with our data. Some of these studies have demonstrated co-localization with endosomes,[38,61,63] lysosomes,[38] mitochondria, or both lysosomes and endoplasmic reticulum.[64]

TEM representative micrographs of cells incubated for 24 hours with CeO$_2$@P2 are presented on Figure 5d. bEnd.3 cells did not show any sign of alteration that might have been induced by particles such as intracellular vacuoles. Moreover, CNPs did not aggregate at the plasma membrane. The lower panels in Figure 5d, which show enlarged views of the micrograph, reveal individual and slightly aggregated particles located only in the compartments attached to the membrane. These compartments are identified as endosomes. Electron microscopy is conventionally used to characterize the shape and size of CNPs; however, only a few used TEM to determine the subcellular location of nanoceria.[24,59,61,62,65] All these studies have shown, as in the present work, a localization of CNPs exclusively in the cytoplasm, and more precisely in vesicles such as endosomes, lysosomes and phagosomes. Studies have shown that CNPs were internalized by cells through an energy-dependent uptake: endosome- and lysosome-involved endocytosis, mainly governed by clathrin-mediated and macropinocytosis pathways.[38,60,64]

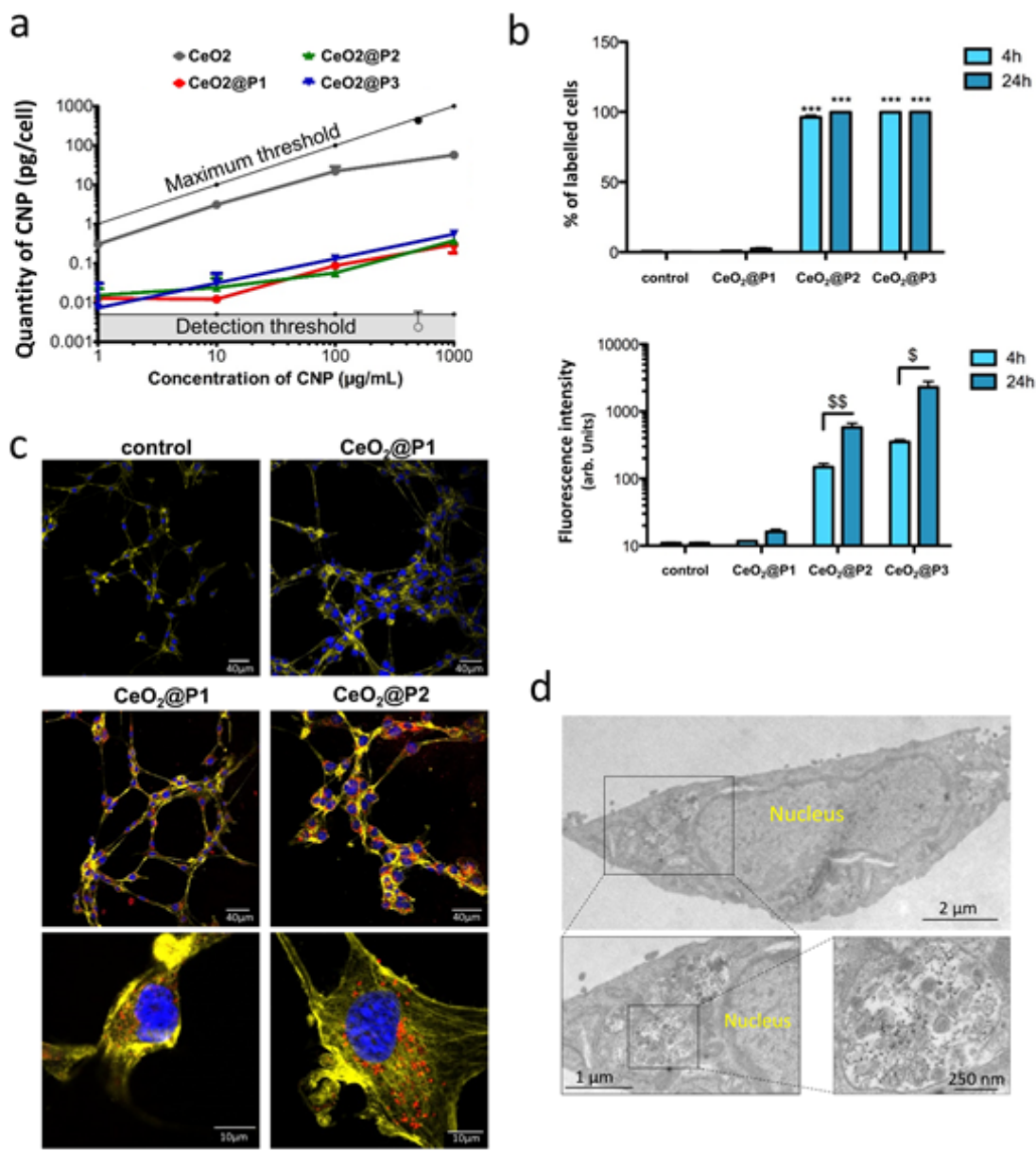

**Figure 5:** *Adsorption, internalization and intracellular location of nanoceria in bEnd.3 cells. (a) ICP-OES nanoceria quantification in b.End3 cells. Bare nanoceria and $CeO_2$@P1, $CeO_2$@P2 and $CeO_2$@P3 were incubated at 1, 10, 100 and 1000 µg mL$^{-1}$ during 24 h. (b) Flow cytometry was performed using cyanine-labelled $CeO_2$@P2 and $CeO_2$@P3 incubated during 4 h or 24 h at 100 µg mL$^{-1}$. Top: percentage of labelled cells, bottom: mean fluorescence intensity per cell. Data are expressed as mean ± SEM, n = 5. *** P < 0.001 versus control, $ P < 0.05 $$ P < 0.01 versus 4 hours. (c) Confocal fluorescence microscopy with cyanine-labelled CNPs (100 µg mL$^{-1}$ during 24 h). bEnd.3 cells were fixed and stained for actin (in yellow) and nuclei (in blue), cyanine-labelled CNPs appear in red. (d) TEM images of bEnd.3 cells incubated with $CeO_2$@P2 (1000 µg mL$^{-1}$ 24 h). Bottom images, which are closer views of the delimited areas on the first and the second pictures, show CNPs enclosed in an endosome near the nucleus.*

### 2.2.3. Antioxidant effect of nanoceria on bEnd.3 cells

*Effect of nanoceria on total intracellular ROS production*

Intracellular ROS production was measured on bEnd.3 cells using the dichlorofluorescein diacetate ($H_2$DCFDA) probe. N-Acetylcysteine (NAC) was used as an antioxidant benchmark. The effect





of nanoceria on ROS production was first assessed on bEnd.3 resting cells (**supplementary Information S5**). The antioxidant NAC significantly reduced the basal levels of intracellular ROS by at least 50% at both incubation times ($P < 0.01$, $P < 0.001$), allowing us to validate the present ROS detection technique. After a 4-hour incubation, only $CeO_2@P1$ at 100 µg $mL^{-1}$ and $CeO_2@P3$ at 1000 µg $mL^{-1}$ reduced intracellular ROS generation by 22% ($P < 0.05$) and 21% ($P < 0.01$) respectively (**supplementary Information S5**). After 24 hours of incubation, none of the CNPs reduced ROS production. The absence of antioxidant effect in resting cells is not a disadvantage since ROS produced in physiological conditions play a role within the cells, particularly in cell signaling.

The effect of nanoceria on bEnd.3 cells' ROS production in oxidative conditions was then examined (Figure 6). Cells were treated with glutamate to mimic the massive release of this neurotransmitter during ischemic stroke that leads to oxidative stress. Glutamate significantly increased ROS levels in bEnd.3 cells at both 4 h (+34%; $P < 0.001$) and 24 h (+44%; $P < 0.05$). The antioxidant NAC significantly reduced glutamate-induced ROS production by about 2-fold ($P < 0.001$) at both incubation times. All nanoceria induced a significant decrease in ROS generation at 4 hours, except for $CeO_2@P3$ at the lowest dose and bare CNPs at 1000 µg $mL^{-1}$. At 24 hours, bare $CeO_2$ did not show any antioxidant effect. $CeO_2@P1$ exhibited an antioxidant effect at 100 µg $mL^{-1}$ ($P < 0.05$), $CeO_2@P2$ at both 100 µg $mL^{-1}$ and 1000 µg $mL^{-1}$ ($P < 0.01$), and $CeO_2@P3$ at 1000 µg $mL^{-1}$ ($P < 0.001$). Taken together, our data demonstrate that our coatings preserve the antioxidant capacities of CNPs in a model relevant to ischemia. First of all, glutamate significantly increased ROS levels in bEnd.3 cells, which is consistent with previous studies performed on bEnd.3 cells[66,67]. Although numerous studies have shown antioxidant effects of CNPs (reduction in ROS production by 30 to 50% at concentrations of 1 to 200 µg $mL^{-1}$),[16,17,21,24,38] only few of them have been performed on endothelial cells.[42,58,60] Additionally, most studies used $H_2O_2$ to induce oxidative stress,[16,24,60] although it may partially react with the probe as well as CNPs in the extracellular medium[68].

## *Effect of nanoceria on mitochondrial ROS production*

We then explored the effect of nanoceria especially on mitochondrial production of superoxide anions using the MitoSOX[TM] Red reagent. In a first step, we evaluated the mitochondrial ROS production after a 4-hour treatment with glutamate (100 mM) or $H_2O_2$ (2 mM) (**Supplementary information S6**). The percentage of MitoSOX[TM] Red reagent-positive cells was not increased by glutamate compared to control cells (**Figure S6**). By contrast, $H_2O_2$ significantly increased the number of labelled cells ($P < 0.01$), in agreement with another study performed on bEnd.3 cells.[66] Based on these results, the effect of CNPs on mitochondrial production of superoxide anions was evaluated in the presence of $H_2O_2$ at 2 mM (Figure 7a). Incubation with $H_2O_2$ induced a 4-fold increase in the fluorescence intensity per cell (112 ± 13 AU, $P < 0.001$ *versus* control cells). Bare CNPs did not induce any change in fluorescence intensity per cell (122 ± 13 AU) while it tended to decrease with coated CNPs: a non-significant 27% decrease for $CeO_2@P1$ (82 ± 15 AU), a 34% reduction close to statistical significance for $CeO_2@P2$ (74 ± 13 AU, P= 0.0848) and a significant 40%-reduction for $CeO_2@P3$ (67 ± 6 AU, $P < 0.05$, Figure 7a). Due to the fast degradation of superoxide anions, our study only focused on the 4-hour treatment. A study conducted on astrocyte showed increased mitochondrial superoxide anion production minutes after glutamate treatment.[69] Although CNPs induced a 32-47% decrease in mitochondrial superoxide anion production, only $CeO_2@P3$ showed a statistically significant effect. Another study demonstrated a significant decrease of only 15% in MitoSox Red levels after addition of CNPs in a hippocampal slice ischemia model. [27]





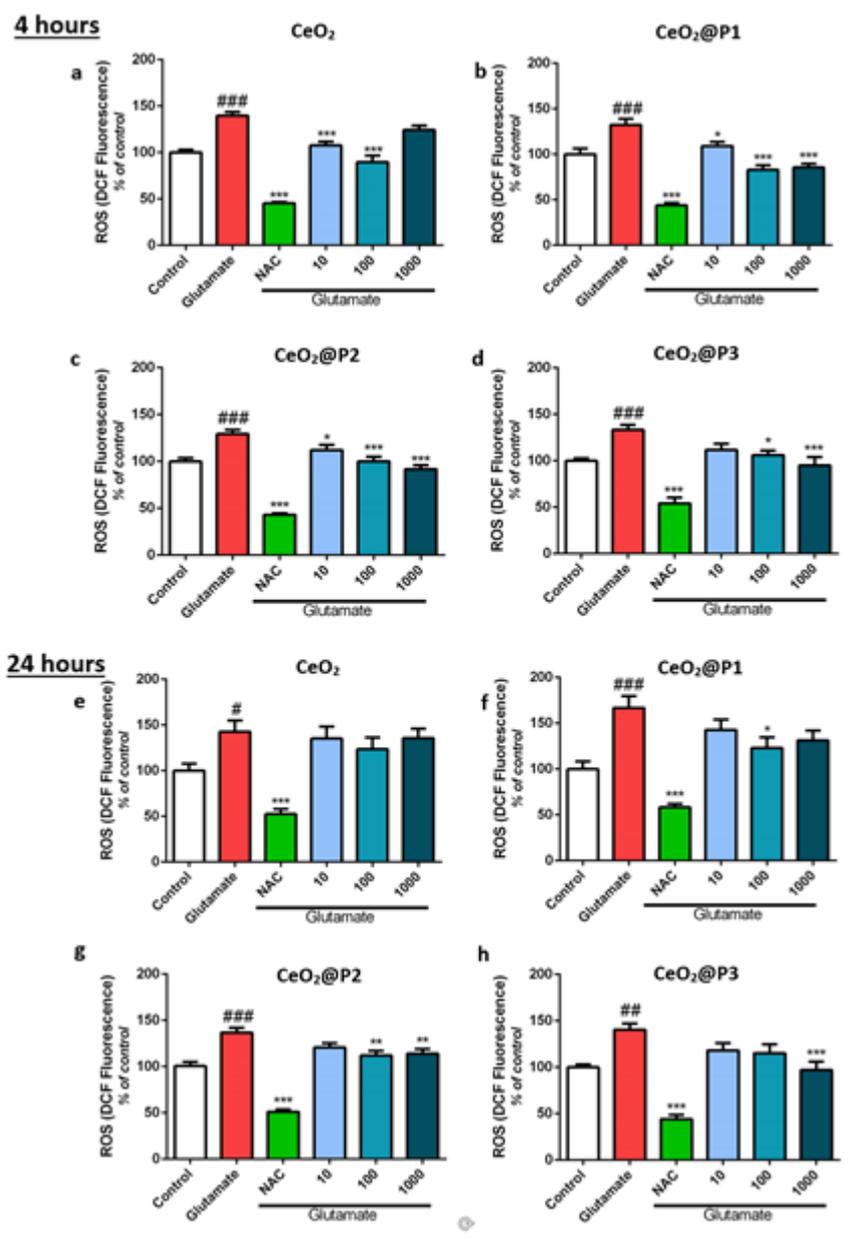

**Figure 6:** *Effect of nanoceria on b.End3 ROS production in oxidative conditions. ROS production was measured with $H_2$-DCF-DA after 4 hours **(a, b, c, d)** and 24 hours **(e, f, g, h)** of incubation. Results were expressed in percentage of fluorescence of control cells. Glutamate (100 mM) was used as pro-oxidant and NAC was used as antioxidant benchmark at 1 mM. Bare nanoceria (a, e), $CeO_2$@P1 **(b, f)**, $CeO_2$@P2 **(c, g)** and $CeO_2$@P3 **(d, h)** were incubated at 10, 100 and 1000 $\mu g$ $mL^{-1}$. Data are expressed as mean ± SEM, n = 8 - 16 at 4 hours and n = 6 - 14 at 24 hours, $^{\#}$ P < 0.05 $^{\#\#}$ P < 0.01, $^{\#\#\#}$ P < 0.001 versus control, * P < 0.05 ** P < 0.01, *** P < 0.001 versus glutamate.*

*Effect on total thiols*

Total thiols have a crucial role as an endogenous antioxidant system. Thiol functions (-SH) maintain the redox state of cells and represent one of the first barriers against ROS which oxidize them, leading to the formation of disulfide bridges (S-S). A decrease in the amount of total thiols is therefore a reflection of oxidative stress. The amount of total thiols on bEnd.3 resting cells was 156 ± 7 $\mu mol/\mu g$ of protein and 72 ± 4 $\mu mol/\mu g$ of protein respectively after 4 hours and 24 hours (Figure 7b, c). Glutamate treatment drastically reduced the amount of total thiols at both 4 hours





(-31%, P < 0.01) and 24 hours (-68%, P<0.001). However, none of the CNPs limited the consumption of total thiols induced by glutamate (Figure 7b, c). Only $CeO_2@P3$, after 24 hours of treatment, showed a non-significant trend towards a decrease in thiols consumption (+21% compared to glutamate alone; Figure 7c). CNPs were unable to limit the consumption of total thiols. Other studies that highlighted a reduction of total thiols's consumption used higher doses of CNPs: 17.2 mg /mL *in vitro*[21] and 60 to 1000 mg/kg *in vivo*[70,71]. In addition, these studies used pretreatments with CNPs, from 48 to 72 hours before induction of oxidative stress[21,70]. This could suggest a better affinity of ROS for thiols and their consumption even before the CNPs have entered the cells.

*Effect on DNA damages*
Among the cellular damages related to oxidative stress, we focused on DNA damages. Guanine is the DNA base most susceptible to oxidative damages, the main modification being its oxidation to 8-hydroxyguanine (8-OHdG).[67] We thus investigated the presence 8-OHdG in bEnd.3 cells after treatment with glutamate, whether or not associated with nanoceria. Glutamate induced a statistically significant increase in DNA oxidation of 149% after a 4-hour incubation (P < 0.001, Figure 7d) and 49% after a 24-hour incubation (P < 0.01, Figure 7e). At 4 hours, bare $CeO_2$, $CeO_2@P1$ and $CeO_2@P3$ induced a significant decrease in glutamate-induced DNA oxidation of 46% (P < 0.001), 26% (P < 0.001) and 23% (P < 0.05) respectively. $CeO_2@P2$ CNPs have no effect on DNA oxidation (Figure 7d). At 24 hours, although a non-significant trend was observed for bare $CeO_2$ (-24%; P= 0.0784) and $CeO_2@P3$ (-15%). Only $CeO_2@P1$ significantly diminished the glutamate-induced DNA oxidation of 41% (P < 0.001; Figure 7e). $CeO_2@P2$ reduced nucleic acid oxidation at both 4 hrs and 24 hrs and only at 24 hrs for $CeO_2P3$. These results were consistent with another study demonstrating a 44% increase in 8-OHdG levels following hypoxia, reduced by 23% by pretreatment by CNPs[72].

In conclusion, *in vitro* experiments have shown that our coatings limit the CNPs' toxicity without altering neither their internalization nor their antioxidant capacities. All coated CNPs substantially display the same overall antioxidant capacities but only $CeO_2@P3$ reduced significantly the production of mitochondrial superoxide anion (Figure 7a) and only $CeO_2@P1$ reduced significantly the DNA oxidation at 24 hours (Figure 7e). In addition, $CeO_2@P3$ carries more amine functions than $CeO_2@P2$, those also being more easily accessible, possibly allowing a greater grafting capacity of targeting peptides. $CeO_2@P3$ and $CeO_2@P1$ (used as a control) have thus been selected for further *in vivo* studies.

## 2.3. *In vivo* study of nanoceria toxicity
### 3.3.1. Biodistribution of nanoceria
The biodistribution of $CeO_2@P3$ – Cy5 (5mg/kg) was followed over one month by measuring fluorescence on the whole mouse and on specific organs using ROIs (Figure 8). The front part acquisitions revealed a stable and maximal fluorescence intensity in the first minutes following the $CeO_2@P3$'s injection and significantly higher than $CeO_2@P1$ (66 ± 14 AU; 66 ± 11 AU and 75 ± 16 AU respectively at 5, 15 and 30 minutes; P <0.001; Figure 8b). Then, the amount of fluorescence slowly decreased and returned to basal level around 14 days post-injection (0.7 ± 0.2 AU). Concerning the back part acquisitions, the fluorescence intensity was 3.5 fold weaker (18 ± 2 AU at 5 minutes) and elimination was faster (Figure 8e). The higher fluorescence level in the front of the mice (Figure 8b, e) suggested the presence of CNPs in spleen, liver (Figure 8c) and bladder (Figure 8d) rather than kidneys (Data not shown).





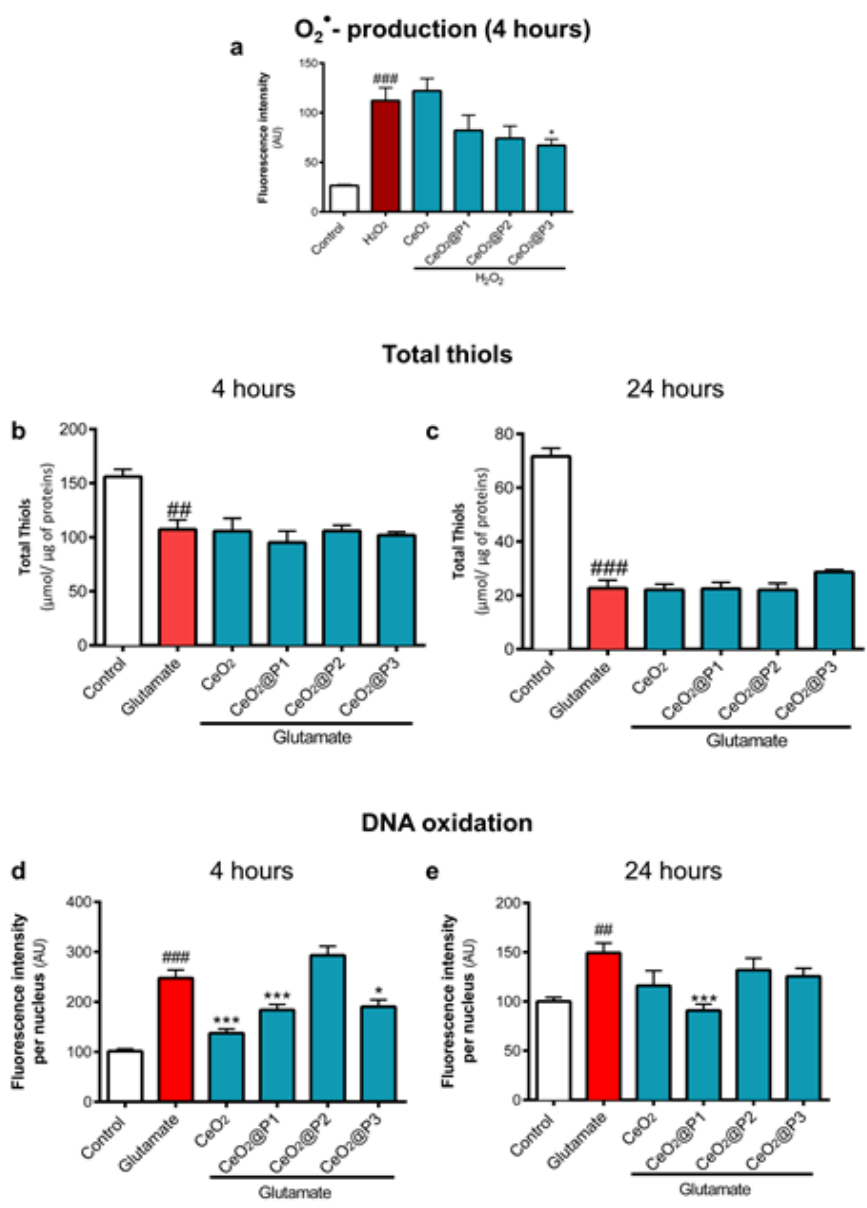

**Figure 7:** *Antioxidant activity of CNPs (100µg/mL⁻¹). Effect of CNPs on mitochondrial production of superoxide anions induced by 4-hour $H_2O_2$ (2 mM) treatment (**a**). The results are expressed in fluorescence intensity per cell in arbitrary units (AU). The data are expressed in mean ± SEM (n = 8). ANOVA and Dunett's test ### P < 0.001 versus control, \* P < 0.05 versus H2O2. Effect of glutamate (100 mM) and CNPs on total thiols amount after 4-hour (**b**) and 24-hour treatments (**c**). The results are expressed as µmol of total thiols per µg of protein. Data are expressed as mean ± SEM (n = 4-8). ANOVA and Dunett's test, ### P < 0.01, ### P < 0.001 versus control. Effect of glutamate (100 mM) and CNPs on DNA oxidation quantified by 8-OHdG measurement after 4-hour (**d**) and 24-hour treatments (**e**). The results are expressed as the percentage of fluorescence of the controls. Data are expressed as mean ± SEM (n = 7). ANOVA and Dunett's test, ### P < 0.001 versus control and \* P < 0.05, \*\*\* P < 0.001 versus glutamate.*

The ROI placed on the liver showed a large and stable amount of CNPs for up to 30 minutes (60 ± 7 AU; 70 ± 8 AU, 58 ± 5 AU, respectively at 5, 15, 30 minutes; P <0.001; Figure 8c), which started to decrease from 60 minutes post-injection (42 ± 5 AU; P <0.001), this reduction being confirmed in the following hours (12 ± 2AU and 8 ± 1AU respectively at 3 hours and 24 hours). Only the ROI drawn on the bladder showed a very important fluorescence peak at 30 minutes (257 ± 87 AU; P





< 0.01; Figure 8d), followed by a rapid decrease to reach a negligible amount from 6 hours post injection.

Twenty-four hours after injections (Figure 8f) $CeO_2$@P3 – Cy5 were mainly found in the liver (33.2 ± 1.6 µg/g; P < 0.001) and spleen (29.8 ± 1.4 µg/g; P < 0.001). A small amount was also found in the kidneys (3.3 ± 0.2 µg/g; P < 0.01) and lungs (3.1 ± 0.2 µg/g; P < 0.05). One month after injections, $CeO_2$@P3 – Cy5 were still mainly found in the spleen (10.1 ± 1.4 µg/g; P < 0,001) and liver (3.2 ± 0.6 µg/g; P < 0.001). A small, non-significant amount of NPCs were found in other organs. In addition, the amount of $CeO_2$@P3 – Cy5 in the liver and spleen decreased significantly compared to 24 hours (P <0.001). Taken together, our data demonstrated $CeO_2$@P3-Cy5's elimination within the first hours after injection. Elimination appeared to be mainly mediated by the liver and spleen with short-term accumulation in these organs. Given the gaussian size distribution of the $CeO_2$@P3-Cy5 measured by DLS (from 10 to 100 nm; data not shown), the smallest NPCs could pass through the kidneys and then the bladder. Most studies were consistent with our data and detected more CNPs in liver and/or spleen compared to bladder or kidneys[18,73–78]. These studies showed an amount of CNPs in the liver 24 hours after the injections from 7µg/g[18,75] to 25µg/g[74]. To note, a study demonstrated a significant concentration of NPC in the bladder in the first hours after the injection without any trace found in the liver and spleen[79].

### 3.3.2. Toxicity of nanoceria

Toxicity of $CeO_2$@P3-Cy5 and of $CeO_2$@P1 used as a control was examined in male Swiss mice. Blood count and anatomo-pathological analysis were performed at both 24 hours and 1 month after nanoceria (5 mg kg$^{-1}$) or PBS intravenous injection. Despite a slight but significant difference in lymphocytes and neutrophils percentages between $CeO_2$@P3-Cy5-treated and control mice (Table III, P < 0.01), blood count values are within the physiological ranges, which indicates a lack of toxicity of our CNPs in the mouse up to 1 month after a single injection at a dose of 5 mg kg$^{-1}$. Our results are in agreement with studies using other CNPs at 0.5 mg kg$^{-1}$ or 5 to 20 mg kg$^{-1}$.[18,80] To note, at higher concentrations (from 30 mg kg$^{-1}$ to 300 mg kg$^{-1}$), it has been reported that CNPs can induce toxicity on blood formula especially on leucocytes and lymphocytes.[76]





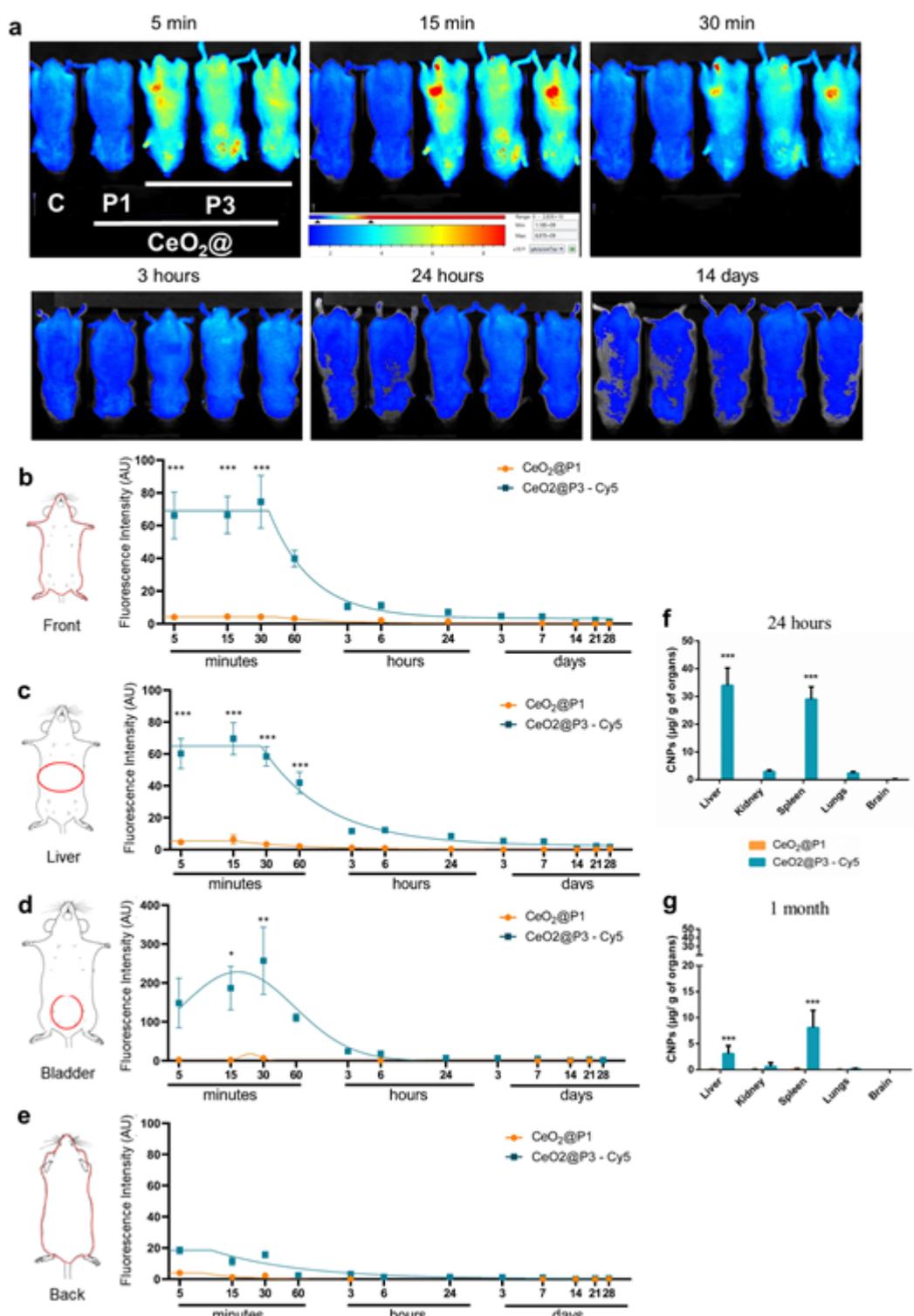

**Figure 8:** *Biodistribution of CeO$_2$@P3 − Cy5 at 5mg/kg in mice. Evolution of fluorescence intensity on frontal acquisitions at different times on control (C), CeO$_2$@P1 (P1) and CeO$_2$@P3 − Cy5 (P3) mice (**a**). Cyanine quantifications after acquisitions of front (**b**), liver (**c**), bladder (**d**) and back (**e**). The fluorescence intensity was expressed in photon / second / cm² / steradian and was calculated from the formula: [(ROI$_{CNP}$ at T$_X$ - ROI$_{CONTROL}$ at T$_X$) / ROI at T$_{5min}$] / ROI$_{STANDAD\ CYANINE}$ at T$_X$. Data are expressed as mean ± SEM (n = 3 for control and CeO$_2$@P1 and n = 9 for CeO$_2$@P3-Cy5). Quantification of cyanine in the organs of mice (Liver, Kidney, Spleen, Lungs, Brain) 24 hours (**f**) and 1 month (**g**) after injection. Data are expressed as mean ± SEM (n = 6). 2-way ANOVA and Sidaks test, * P < 0.05, ** P < 0.01 and *** P < 0.001 versus CeO$_2$@P1 and $$$ P < 0.001 versus 24 hours.*

**Table III** – *Blood count of mice treated with PBS (control), CeO2@P1 and CeO2@P3-Cy5. Nanoceria were administrated intravenously (5 mg kg$^{-1}$, 5 mL kg$^{-1}$ in PBS 0.1M) and blood count was assessed 24 hours and 1 month after injection. **: P < 0.01 versus control.*

| | 24 hours | | | 1 month | | |
|---|---|---|---|---|---|---|
| | Control | CeO2@P1 | CeO2@P3-Cy5 | Control | CeO2@P1 | CeO2@P3-Cy5 |
| Leukocytes (10³/mm³) | 4.6 ± 0.7 | 4.1 ± 0.5 | 4.5 ± 0.5 | 3.8 ± 0.7 | 2.6 ± 0.4 | 3.9 ± 0.5 |
| Lymphocytes (%) | 69.6 ± 2.1 | 62.3 ± 3.0 | 57.3 ± 1.9** | 60.9 ± 4.5 | 66.0 ± 2.5 | 69.8 ± 2.9 |
| Monocytes (%) | 5.9 ± 0.4 | 5.6 ± 0.3 | 5.9 ± 0.4 | 6.6 ± 0.5 | 6.5 ± 0.4 | 5.7 ± 0.3 |
| Neutrophils (%) | 20.0 ± 2.0 | 27.2 ± 2.5 | 31.0 ± 1.9** | 26.9 ± 4.0 | 19.6 ± 2.5 | 19.5 ± 2.7 |
| Eosinophils (%) | 1.1 ± 0.6 | 1.4 ± 0.3 | 1.7 ± 0.5 | 2.0 ± 0.5 | 3.9 ± 1.2 | 1.0 ± 0.5 |
| Basophils (%) | 1.7 ± 0.2 | 1.7 ± 0.1 | 1.4 ± 0.3 | 1.6 ± 0.3 | 1.8 ± 0.2 | 1.9 ± 0.3 |
| Erythrocytes (10⁹/mm³) | 8.3 ± 0.3 | 8.4 ± 0.4 | 8.3 ± 0.2 | 9.4 ± 0.2 | 8.9 ± 0.2 | 9.0 ± 0.2 |
| Hematocrit (%) | 41.5 ± 2.1 | 43.1 ± 2.2 | 42.4 ± 1.0 | 45.6 ± 1.2 | 42.7 ± 0.9 | 42.2 ± 0.9 |
| Hemoglobin (g/dL) | 10.6 ± 0.5 | 11.6 ± 0.3 | 11.1 ± 0.3 | 12.6 ± 0.4 | 12.0 ± 0.3 | 12.2 ± 0.3 |
| Thrombocytes (10³/mm³) | 1809 ± 93 | 1786 ± 114 | 1719 ± 73 | 1783 ± 92 | 1835 ± 69 | 1724± 87 |

Concerning anatomo-pathological analysis, regarding the spleen, CNPs did not induce any lesion, and the distribution and the size of follicles was unaffected by the treatment neither at 24 hours (data not shown) nor at 1 month post injection (Figure 9). The homogenous distribution of the white and red pulp highlights the representativeness of the cell types and the immunocompetence of the observed spleens. For the liver, no lesion or fibrosis was induced by CNPs injections. Some regions of regeneration were observed but no difference was observed compared to control mice. In kidneys, no CNPs-induced lesions were observed, especially in glomeruli and Bowman's capsules. Regarding the lungs, no fibrosis was observed following the injection of CNPs. Some hemorrhagic infiltrates were detected especially in the alveolar ducts. However, these infiltrates were also detected in control mice and would therefore seem to be a post-mortem artefact. Finally, no lesions were observed in the brain after CNPs injection. Most studies reported an absence of CNPs toxicity. Similar results have been reported 24 hours to 3 months after injection of CNPs at doses of 0.5 to 100 mg kg$^{-1}$.[18,74,75] The only studies that demonstrated toxicity after injection of CNPs used very high doses: greater than 250 mg kg$^{-1}$ by intravenous injection[76] and 641 mg m$^{-3}$ by inhalation[81]

Overall, these data demonstrate the lack of toxicity of CNPs both 24 hours and 1 month after a single injection of coated CNPs at 5 mg kg$^{-1}$. No abnormal blood count, lesion, necrosis, fibrosis, cellular regeneration, significant hemorrhagic or immune infiltrate was observed. The lack of *in vivo* toxicity of our coated CNPs allows us to plan the next study of their antioxidant effect in an experimental model of cerebral ischemia to assess their potential for therapeutic use.

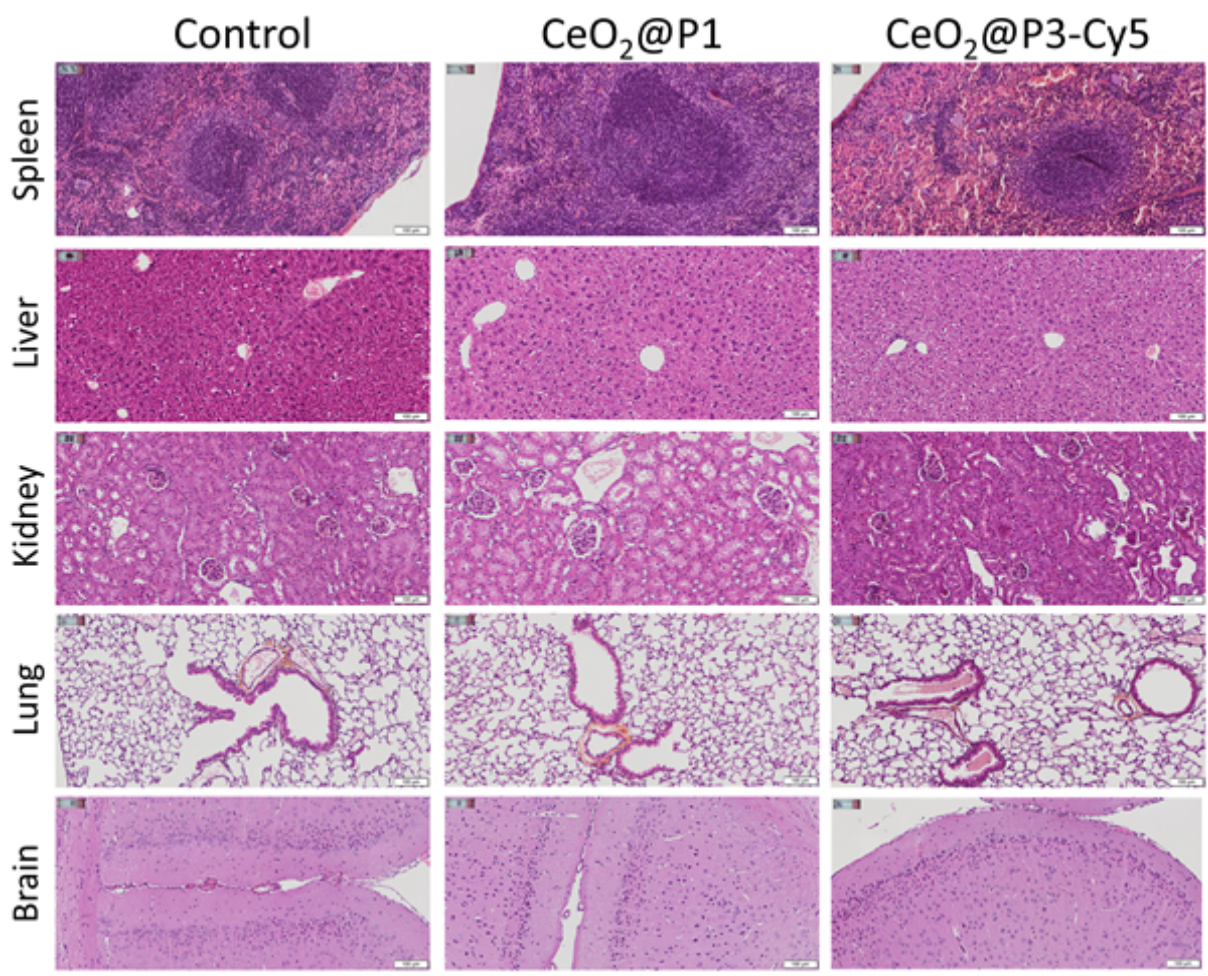

**Figure 9**: *Anatomo-pathological analysis of mice treated with PBS (control), CeO$_2$@P1 or CeO$_2$@P3-Cy5 one month after injection. PBS and nanoceria were administrated intravenously (5 mg kg$^{-1}$, 5 mL kg$^{-1}$ in PBS 0.1M).*

Overall, these data demonstrate the lack of toxicity of CNPs both 24 hours and 1 month after a single injection of coated CNPs at 5 mg kg$^{-1}$. No abnormal blood count, lesion, necrosis, fibrosis, cellular regeneration, significant hemorrhagic or immune infiltrate was observed. The lack of *in vivo* toxicity of our coated CNPs allows us to plan the next study of their antioxidant effect in an experimental model of cerebral ischemia to assess their potential for therapeutic use.

## 3. Conclusion

To conclude, we formulated cerium oxide nanoparticles coated with statistical multiphospho-nate-PEG copolymers as enzyme-mimicking catalysts for the decomposition of reactive oxygen species. The polymers examined have a dual functionality, one for protection against protein adsorption in the form of a PEGylated corona and one for targeting thanks to terminal amine groups allowing further covalent binding. These polymers impart remarkable colloidal stability to metal oxide nanoparticles in biological media. Keeping the CNP core identical and changing the nature of the coating, we studied the impact of polymers on the catalase- and superoxide dismutase-like catalytic activities of bare and coated cerium oxide. We observed that the multi-PEG coatings did not affect the superoxide dismutase-like and slightly impair the catalase-like effect of nanoceria, a result that confirms the benefit of having phosphonic acids as anchoring groups at the particle surface. *In vitro* assays performed on murine cerebral endothelial cells showed that these coated CNPs do not exert any toxicity while interacting with cells, as demonstrated by ICP-OES and flow cytometry measurements. Confocal microscopy and TEM studies notably indicated a perinuclear cytoplasmic CNPs localization, more precisely in the endosomes. Coated CNPs were able to reduce intracellular glutamate-induced ROS production, showing that the coating does not decrease the antioxidant effect. While all coated CNPs exert substantially the same antioxidant effects, evaluation of mitochondrial ROS production and DNA oxidative damages allow identifying $CeO_2$@P3 as the most effective nanoparticles. Finally, IV administration of $CeO_2$@P3 in mice highlighted a lack of toxicity on blood count and anatomo-pathological analysis up to 1 month as well as an elimination mainly via the liver and the spleen. The present work therefore enables us to identify biocompatible CNPs with very interesting antioxidant effects that can be used for the treatment of cerebral ischemia. The continuation of this work will therefore consist in evaluating their efficacy in models of stroke, for subsequent clinical application. It will also be of particular interest to investigate whether the grafting of targeting agents onto these particles would make it possible to concentrate their antioxidant effects in the affected areas, thus increasing their beneficial effects.

# 4. Materials and Methods

## 4.1. Nanoceria coating and characterization

### Materials

The cerium oxide nanoparticles aqueous dispersion (200 g $L^{-1}$, pH 1.5) was synthesized by Rhodia (Centre de Recherche d'Aubervilliers, Aubervilliers, France).[49,50] Polymers were synthesized and provided by SPECIFIC POLYMERS (Castries, France). Poly(ethylene glycol) mono-functional Vinylether (Vinyl-PEG, CAS: 79−41−4) was supplied by Ineos and used as received. Methacryloyl chloride (CAS: 920-46-7, Sigma Aldrich) was distilled before use. Dimethyl(methacryoyloxy)methyl phosphonate ($MPh_e$, SP-41-003, CAS: 86242−61−7) monomers were produced by SPECIFIC POLYMERS. Cysteamine hydrochloride (CAS:156-57-0), Di-tert-butyl dicarbonate ($Boc_2O$, CAS: 24424-99-5), 2,2′-Azobis(isobutyronitrile) (AIBN; CAS: 78−67−1) were supplied by Sigma-Aldrich and used as received.

### Transmission Electron Microscopy (TEM)

Micrographs were taken with a Tecnai 12 TEM operating at 80 kV equipped with a 1K×1K Keen View camera. Nanoceria dispersions were deposited on ultrathin carbon type-A 400 mesh copper grids (Ted Pella, Inc.). Micrographs were analyzed using ImageJ software for 200 particles.

### X-ray Photoelectron Spectroscopy (XPS)

XPS data was collected using an Omicron Argus X-ray photoelectron spectrometer using a monochromatic $AlK_\alpha$ (1486.6 eV) radiation source with a 300 W electron beam power. The emission of photoelectrons from the sample was analyzed at a takeoff angle of 45° under ultra-high vacuum conditions ($10^{-8}$ Pa). The spectra were collected at pass energy of 100 eV for the survey scan and 20 eV for the high-resolution scans. The XPS spectra were fitted using the software XPSPEAK41, applying a fixed Gaussian/Lorentzian ratio for peaks of the same spectrum and constraining the full-width at half-maximum (FWHM) of each doublet to be equal. The fraction of $Ce^{3+}$ was determined through the ratio between the integrated intensities of the four peaks corresponding to $Ce^{3+}$ divided by the integrated intensities of all ten peak.[13]

*Coating cerium oxide nanoparticles*

Cerium oxide nanoparticles were coated with phosphonic PEG copolymers using a formulation pathway described earlier.[48,51] In brief, particle and polymers dispersions were prepared in the same conditions of pH (pH 1.5) and concentration ($c$ = 2 g $L^{-1}$) and mixed at different volume ratios. The CNP dispersion was added dropwise to the polymer solution under magnetic stirring keeping the mixing volume ratio at $X_C/5$ where $X_C$ denotes the critical mixing ratio (nanoparticle over polymer) above which the CNPs are partially coated and precipitate at physiological pH.[45] Working at $X_C/5$ insures that polymers are in excess during adsorption. After increasing their pH to 8 by addition of $NH_4OH$, the dispersions were centrifuged at 4000 rpm using Merck centrifuge filters (pore 100000 g $mol^{-1}$) to remove the polymer excess and further concentrated to 20 g $L^{-1}$. The hydrodynamic diameter $D_H$, electrophoretic mobility and zeta potential were obtained using dynamic light scattering (DLS) and electrokinetic measurements (NanoZS Zetasizer spectrometer, Malvern Instruments). The hydrodynamic diameters provided here are the second coefficients in the cumulant analysis derived from the Stokes-Einstein relation $D_H = k_B T/3\pi\eta D_C$ where $k_B$ is the Boltzmann constant, $T$ the temperature, $\eta$ the solvent viscosity and $D_C$ the average diffusion coefficient. Measurements were performed in triplicate at 25 °C after an equilibration time of 120 s. A UV-visible spectrometer (SmartSpecPlus from BioRad) was used to measure the absorbance of polymer coated nanoceria aqueous dispersions. Absorbance data were used to determine the nanoparticle concentration for each batch by means of Beer-Lambert law.

*SOD mimetic activity assay*

The superoxide dismutase kit assay was purchased from Sigma-Aldrich (Lyon, France). The catalytic activity of nanoceria in the dismutation of superoxide radical anion was assessed by a colorimetric assay using UV-Vis spectroscopy (Kit #19160-1KTF). Briefly, 20 μL of a nanoceria dispersion in Tris-Cl buffer pH 7.5 was added to a well of a 96-well plate and mixed with 200 μL of WST-1 (2-(4-Iodophenyl)-3-(4-nitrophenyl)-5-(2,4-disulfophenyl)-2H-tetrazolium, monosodium salt). The reaction was initiated with the addition of 20 μL of xanthine oxidase solution, prepared by mixing 5 μL of the enzyme in 2.5 mL of a dilution buffer provided. After incubating plate at 37°C for 20 min, the absorbance at 450 nm was measured using a microplate reader (EnSpire Multimode Plate Reader, Perkin Elmer). Nanoceria final concentration ranged from 1 to 1000 μg $mL^{-1}$.

*CAT mimetic activity assay*

The Amplex® Red catalase assay kit was obtained from Thermo Scientific (Illkirch, France). The catalytic activity of nanoceria in the disproportionation of $H_2O_2$ was assessed by spectrofluorimetry using the Amplex-Red reagent assay (Cat # A22180). Briefly, 25 μL of each nanoceria dispersions in Tris-Cl buffer pH 7.5 was added to a well of a 96-well plate and mixed with 25 μL of a $H_2O_2$ solution, to obtain an $H_2O_2$ concentration of 5 μM in each well. Then, 50 μL Amplex Red reagent/HRP working solution was added and reactions pre-incubated for 5 minutes. Amplex Red (10-acetyl-3,7-dihydroxyphenoxazine) reaction with $H_2O_2$ catalyzed by horseradish peroxide (HRP) produces the fluorescent molecule resorufin (excitation at 571 nm and emission at 585 nm). The fluorescence was measured after incubating for 30 min with protection from light. The final $H_2O_2$ concentration in each well was 5 μM, whereas those of nanoceria ranged from 1 to 1000 μg $mL^{-1}$.

**4.2. *In vitro* experiments**

*Cell Culture*

Immortalized mouse brain endothelial cells, bEnd.3 (ATCC® CRL-2299™, Manassas, Virginia, USA) purchased from Sigma (Sigma-Aldrich, Saint Quentin Fallavier, France) were cultivated in Dulbecco's modified Eagle's medium (DMEM) Glutamax (Gibco, 31966-021), supplemented with 10% foetal bovine serum (FBS, Dutscher S1810-500), 100 $U.mL^{-1}$ penicillin and 100 µg $mL^{-1}$ streptomycin (Gibco, 15140122) in a humidified 5% $CO_2$ incubator at 37 °C. Cell line passages < 30 were used for all experiments. To evaluate mortality, metabolic activity and measure ROS, bEnd.3 cells were seeded in 96 well plates at $5 \times 10^4$ cells per well (≈ 100 000 cells $cm^{-2}$); for mitochondrial ROS detection, cells were seeded into 24-well plates at a density of 200 000 cells per well (≈ 100 000 cells $cm^{-2}$). Twenty-four hours after, treatments were applied (Glutamate 100 mM, NAC 1 mM, $H_2O_2$ 2 mM and CNPs at 10, 100, and 1000 µg $mL^{-1}$) diluted in DMEM without FBS and without phenol red (Gibco, 21063-09), and incubated with cells during 4 h or 24 h.

*Trypan Blue count*
The cells were detached from the wells by trypsin-EDTA (0.25%), inactivated by complete medium, and centrifuged at 100 g for 5 minutes. Cells were then taken up in 100 µL of PBS with 1% of Bovine Serum Albumin (BSA) and stained with trypan blue 0.2% (Sigma, T8154). Blue and white cells were then counted in Kova slides (Dutscher, 050126) with a minimum of 150 cells. The percentage of mortality was calculated: $100 \times$(Blue Cells / (White + Blue cells)).

*MTT assay*
A solution of 5 mg $mL^{-1}$ of 3-(4,5-Dimethyl-2-thiazolyl)-2,5-diphenyl-2H-tetrazolium bromide (MTT, Sigma, S12045) was prepared, and 10 µL were added to each well (final concentration at 0.5 mg $mL^{-1}$) and incubated for 3h. MTT was reduced to formazan by mitochondrial succinate dehydrogenase within viable cells. The wells were emptied, and formazan was dissolved in DMSO (23.488.294, Prolabo, Fontenay-sous-bois, France). Optical Density was read at a wavelength of 570 nm and nonspecific absorbance at 690nm was removed. Results were expressed in percentage of absorbance of control cells.

*Inductively Coupled Plasma - Optical Emission Spectrometry (ICP-OES)*
Two million bEnd.3 cells ($2 \times 10^6$) were incubated into 6-well plates (≈ 100 000 cells $cm^{-2}$) for 24 h with CNPs dispersed in DMEM (1, 10, 100 or 1000 µg $mL^{-1}$), combined or not with glutamate 100 mM. Cells were washed with PBS and digested with concentrated nitric acid ($HNO_3$, Prolabo, 20.425.297) and hydrogen peroxide (10 M). After evaporation the dry samples were diluted with HNO3 2%. For cerium detection, ICP-OES experiments were performed at the Institut Physique du Globe de Paris (IPGP) using an iCAP6200 Thermofisher spectrometer. The measurements were performed in triplicate at radiation wavelengths 393.11 Å and 404.08 Å with an uncertainty better than 5%.

*Cyanine labelling*
The primary amine groups of $CeO_2@P2$ and $CeO_2@P3$ were quantified by spectrofluorometry (**Supplementary Information S4**). $CeO_2@P2$ and $CeO_2@P3$ were labelled with Cyanine 5 (Lumiprobe, #23320) in PBS buffer adjusted to pH 8 (2 hours at room temperature while stirring continuously). Particles were washed five times with PBS on an exclusion column (30 kDa) to remove the free Cyanine. The same experimental procedure was carried out with the amine-free particle $CeO_2@P1$ as a control, to validate the washing procedure.

*Flow cytometry*

Adsorption/internalization was evaluated by flow cytometry on bEnd.3 cells cultured in 24 well plates ($2 \times 10^5$ cells per well $\approx$ 100 000 cells $cm^{-2}$) in DMEM Glutamax during 24 h. Cyanine 5 labelled $CeO_2$@P2 and $CeO_2$@P3 (100 µg $mL^{-1}$ in DMEM without FBS and without phenol red, Gibco, 21063-09) were incubated during 4 h or 24 h. Cells were washed 3 times with PBS, detached from the wells by trypsin, inactivated by complete medium, and centrifuged at 100g for 5 minutes. The cells were taken up in PBS - BSA - 1% (ID Bio, 1000-70) – EDTA 200nM (Sigma, ED4SS) and cellular uptake was quantified by flow cytometry (Millipore Guava Easy Cyte Ds200, excitation 640 nm, emission 660nm).

*Confocal fluorescence microscopy*

For microscopy studies, the cells were cultured in 24 well plates on coverslips (Knittel Glass, Brunswick, Germany) as described above. $CeO_2$@P1, $CeO_2$@P2-Cy5 and $CeO_2$@P3-Cy5 were incubated during 4 and 24 hours at 100 µg $mL^{-1}$. Cells were then washed 3 times with PBS, fixed with 4% PFA (ACROS, 169650025), added with 0.5% Triton X100 (Sigma, 516813) for permeabilization, during 12 minutes and washed again 3 times. Actin was labelled by phalloidin A488 as described by the supplier (Thermofisher, A12379) and 3 washes with PBS were carried out. The nuclei were labelled by DAPI at 0.1 µg $mL^{-1}$ for 5 minutes (Merck Millipore, 268298) followed by 3 washes with PBS. Finally, the coverslips were mounted with mowiol (Mowiol Sigma 81381, Glycerol Merck 4094 ratio 1:2.5) on glass microscope slides (Roth, Superfrost 2109). Slides were observed with a Leica confocal microscope (Leica Microsystems, DMI8).

*Transmission Electron Microscopy (TEM) on bEnd.3 cells*

bEnd.3 cells sere seeded in 6-wells plates at a density of 1 000 000 cells per well in DMEM medium and cultured for 24 hours. Cells were treated by $CeO_2$@P1 and $CeO_2$@P2 at 1000 µg $mL^{-1}$ in DMEM medium without FBS and without phenol red for 24 hours. Cells were washed with PBS and fixed with 1% glutaraldehyde and 2.5% PFA at 4°C for 48 hours. Cells were scraped and included in 2% gelatin. The pellets were recovered and centrifugated at 250g for 1 minute, followed by a second fixation with 1% osmium tetroxide (OsO4) and potassium ferrate (FeK) for 1 hour at room temperature. An inclusion in 1% agarose was performed and cut into cubes. The dehydration was carried out by successive ethanol baths: 30% (15 min); 50% (15 min), 70% (15 min), 90% (15 min) 100% (2 x 15 min), and 100% ethanol with propylene oxide (2 x 15 min). The impregnation was performed in propylene oxide / Epoxy resin (Epon) at ratio 1:1 overnight followed by 2 baths of Epon for 1 hour. The inclusion was followed by polymerization at 60°C for 18 hours. TEM was performed on 70 nm thick microtome cell sections. The micrographs were made on a Tecnai 12 TEM of 80kV equipped with a camera 1K x 1K KeenView.

*Intracellular ROS detection*

At the end of the treatment, bEnd.3 cells were incubated with the dichloro-fluorescein diacetate probe ($H_2$DCF-DA, Sigma D6883) at 20 µM during 30 minutes at 37°C in the dark. This lipophilic dye passively diffuses through cellular membranes and is cleaved by intracellular esterase to form $H_2$DCF. When reacting with ROS, $H_2$DCF gives fluorescent DCF. Fluorescence was measured with TECAN fluorimeter (Infinite F200 Pro) with an excitation wavelength of 485 nm and an emission wavelength of 535 nm.

*Mitochondrial ROS detection*

The MitoSOX$^{TM}$ Red reagent is a compound that penetrates living cells and selectively targets mitochondria. After incubation of the different treatments, the cells were washed three times with PBS and detached from the wells by trypsin, inactivated by complete medium without phenol red, centrifuged for 5 minutes at 100 g and the supernatant was removed. The MitoSOX™ Red reagent (Thermofisher, M36008) diluted in PBS (5µM) was added in each tube with PBS - BSA 1g L$^{-1}$ (ID Bio, 1000-70) - EDTA (200nM - Sigma, ED4SS) at ratio 1:4 and incubated for 20 minutes. Cellular uptake was quantified with FACS (Millipore Guava Easy Cyte Ds200) at excitation wavelengths of 488nm and emission wavelengths of 580nm.

*Quantification of total thiols*
After treatments, bEnd.3 cells were washed by PBS. Three freezing – thawing cycle were carried out (-80°C to 37°C). Standard of reduced glutathione was performed from 0 to 50µM. Thiol Green Indicator 2X solution (156049, Abcam, Paris, France) was prepared and diluted at ½ in each well. Fluorescence intensity was measured with TECAN fluorimeter (Infinite F200 Pro) at T0 and T20 minutes with an excitation wavelength of 485 nm and an emission wavelength of 535 nm. The results were normalized by the protein concentration. Briefly, the proteins were precipitated by trichloroacetic acid (Sigma T4885). The lysates were collected and centrifuged at 10,000g for 10 minutes. The protein pellet was suspended in 60µL of 1M NaOH and then a BCA assay was performed.

*Quantification of DNA damages (8-OHdG)*
DNA oxidative damages were studied by investigating the presence of 8-hydroxyguanine (8-OHdG) in bEnd.3 cells. Cells were seeded in 24 well plates on coverslips (Knittel Glass, Brunswick, Allemagne) at a density of 200 000 cells/well in DMEM medium and cultured for 24 hours as described above. The cells were treated with DMEM (control cells), glutamate (100mM) alone or associated with CNPs (CeO$_2$@P1, CeO$_2$@P2-Cy5 and CeO$_2$@P3-Cy5 at 100 µg mL$^{-1}$) in DMEM medium free of FBS and phenol red (Gibco, 21063-09) for 4 hours. Cells were washed with PBS, fixed with 4% PFA (169650025, ACROS, Fisher Scientific) for 12 minutes. Non-specific sites were blocked with PBS - goat serum 5% (Sigma, G9023 - Lot SCBL5146V) - triton 0.1% (Sigma, 516813) for 30 minutes. Mouse anti-8-OHdG antibody (SantaCruz, sc66036 Clone 15A3, 1/100) was incubated overnight at 4°C. The AlexaFluor anti-mouse A555 (Invitrogen, A11003) was incubated (1/500) 2 hours at room temperature. DAPI was incubated for 5 minutes (0.1 µg mL$^{-1}$ in PBS, Merck Millipore, 268298). The slides were mounted with mowiol (Mowiol Sigma 81381, Merck Glycerol 4094 ratio 1:2.5) on microscopy slides (Roth, Superfrost 2109). The images were taken with an epifluorescence microscope (Zeiss Axiophot, Marly le Roi, France) at a wavelength of 550nm. The analyses were carried out using Image J software. Fluorescence intensity was quantified at the nucleus level using DAPI labeling.

### 4.3. *In vivo* study
All experiments were carried out in accordance with the European Community Council Directive of September 22, 2010 (2010/63/UE) and the French regulations regarding the protection of animals used for experimental and other scientific purposes (D2013-118), with the approval of both the Paris Descartes University Animal Ethics Committee (Comité d'Ethique pour l'expérimentation animale de Paris Descartes, CEEA 34) and the Ministry of Education and Research (registered number APAFIS Project #13638).
Six-week old male Swiss mice (28–32 g, Janvier Labs, Le Genest-Saint-Isle, France) were housed under standard conditions with a 12 h light/dark cycle and allowed access to food and water ad libitum. Mice were injected with PBS (control mice), CeO$_2$@P1 or CeO$_2$@P3-Cy5 intravenously (tail vein, 5 mL kg$^{-1}$ in PBS 0.1M).

*Biodistribution*

Animals were anesthetized with isoflurane (2% in an air-oxygen mixture) and placed in a Photon Imager (Biospace Lab, Nesles-la-Vallée, France). Acquisitions at early times were carried out for the animals killed at 24 hours: 5, 15, 30 minutes and 1, 3 and 6 hours after injection. Concerning the group killed at 1 month, acquisitions were performed at later stages: 1, 2, 3, 7, 14, 21 and 28 days after injection. The fluorescence intensity was expressed in photon / second / cm² / steradian. Regions of interest (ROI) were used to measure fluorescence intensity in specific organs. The fluorescence intensity from ROIs was calculated from the formula:

$[(ROI_{CNP} \text{ at } T_X - ROI_{CONTROL} \text{ at } T_X) / ROI \text{ at } T_{5min}] / ROI_{STANDAD\ CYANINE} \text{ at } T_X$

Quantification of cyanine was carried out in different organs at 24 hours and 1 month after the injections. Organs were thawed, weighed and transferred to 5 and 7mL tubes containing ceramic beads (P000918 and P000935, Precellys Lysing Kit, Ozyme, Saint-Cyr-l'École, France). Organs were mechanically ground by two cycles 6000 RPM during 30 seconds in PBS / Triton 0.1% (02520-300R-D000, Precellys evolution, Bertin Technologie, Montigny-le-Bretonneux, France). PBS / 0.1% triton volume was adjusted as function of the weight and density of each organ. Lysate were centrifuged at 10,000g for 10 minutes. Cyanine fluorescence was measured with TECAN fluorimeter (Infinite F200 Pro) with an excitation wavelength of 620nm and emission wavelengths of 670nm.

*Toxicity*

An intracardiac blood sample (1 mL) was collected in eppendorf tubes containing 5μl of 0.5 M EDTA (Sigma, 03660) and used for blood count (NFS, MS95, Melet Schloesing Laboratories, Osny, France). Immediately after, animals were decapitated, the organs were removed and post-fixed with paraformaldehyde 4% (Euromedex, 16%, 15700, Souffelweyersheim, France) for 24 hours for histological studies. Histology was performed on 5 organs: liver, kidneys, spleen, lungs and brain. The lungs were inflated with PFA with the help of a syringe and a flexible catheter inserted into the trachea. Inclusion, sections, staining were performed by the PETRA platform (Experimental and Translational Pathology) at the Gustave Roussy Institute (Villejuif, France). The samples were dehydrated using 70% and 100% alcohol baths and with isopropanol and then incorporated into paraffin with LOGOS to dehydrate and clarify the samples (MM, France). Sections of 4 μm were made with a microtome and dried at 37°C. HES staining was performed and then mounting medium (Pertex) was used to glue a coverslip on the tissue section. Each section was observed using a Zeiss Axiophot microscope and the photos were taken using Cameware.

### 4.4. Statistical analysis

Data are presented as mean ± standard error of the mean (SEM). Statistical analyses were performed using GraphPad Prism 5. Data were analyzed using one-way analysis of variance (ANOVA) followed by *post-hoc* Dunnett-test when the data were compared to the control condition, or the Student's t-test with Bonferroni correction when several groups were compared with each other's. Differences were considered significant when $P < 0.05$.

# Supplementary Information

The Supporting Information is available free of charge on Advanced Healthcare Materials Publications website at DOI: 10.1021/advancedhealthcarematerials /xxxxxx. S1: X-ray photoelectron spectrometry (XPS) Ce 3d spectra – S2: Copolymers synthesis and characterization – S3: Quantification of nanoceria primary amine groups – S4: Effect of nanoceria on ROS production on resting bEnd.3 cells – S5: Effect of glutamate and $H_2O_2$ on mitochondrial ROS production

# Acknowledgments

We thank Jérôme Fresnais, Evdokia Oikonomou, Fanny Mousseau, Chloé Puisney, Eugénie Mussard for fruitful discussions. We thank the Cellular and Molecular Imaging facility (INSERM US25, CNRS UMS 3612) of the Faculty of Pharmacy of Paris, University of Paris, France. We thank the "Histo-cytopathologie" platform of Gustave Roussy Institute and especially Olivia Bawa and Dr Paule Opolon for the histological studies. ANR (Agence Nationale de la Recherche) and CGI (Commissariat à l'Investissement d'Avenir) are gratefully acknowledged for their financial support of this work through Labex SEAM (Science and Engineering for Advanced Materials and devices) ANR 11 LABX 086, ANR 11 IDEX 05 02. We acknowledge the ImagoSeine facility (Jacques Monod Institute, Paris, France), and the France BioImaging infrastructure supported by the French National Research Agency (ANR-10-INSB-04, « Investments for the future ») for the transmission electron microscopy support. Parts of this work were supported by IPGP multidisciplinary program PARI, and by Paris–IdF region SESAME Grant no. 12015908. This research was supported in part by the Agence Nationale de la Recherche under the contract ANR-13-BS08-0015 (PANORAMA), ANR-12-CHEX-0011 (PULMONANO), ANR-15-CE18-0024-01 (ICONS), ANR-17-CE09-0017 (AlveolusMimics) and by Solvay.

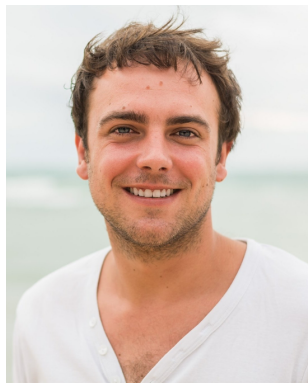

Geoffroy Goujon obtained his master's degree in biology in 2016. During the first year of his master's he worked on iPS cells as a new model for studying ALS (Brain and Spinal Cord Institute, Paris). Then he worked on the impact of varenicline on early hyperactivity related to Alzheimer's disease (Pasteur Institute, Paris). He completed his PhD of Pharmacology at Paris Descartes University in 2019 where he worked on the therapeutic potential of innovative cerium oxide nanoparticles for cerebral ischemia. He is currently working on angiogenesis defects related to HHT disease as a post-doctoral researcher at LUMC (Leiden, Netherlands).

**Table of contents**

Direct and indirect antioxidant effects induce by Cerium Oxide Nanoparticles (CNPs) in endothelial cells.

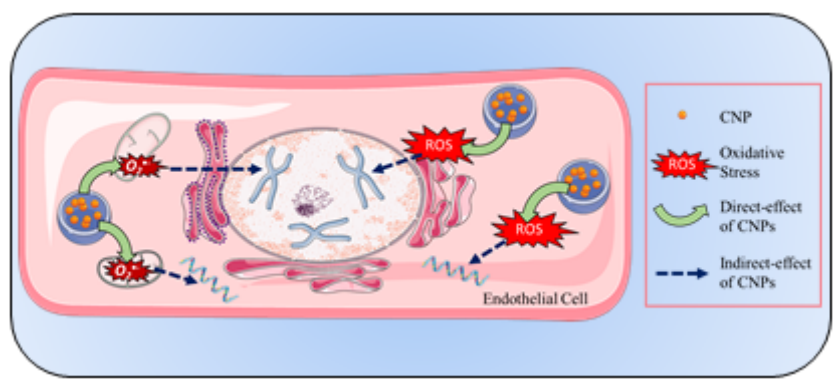

## Supporting Information

**Antioxidant activity and toxicity study of cerium oxide nano-particles stabilized with innovative functional copolymers**

*Geoffroy Goujon, Victor Baldim, Caroline Roques, Nicolas Bia, Johanne Seguin, Bruno Palmier Alain Graillot, Cédric Loubat, Nathalie Mignet, Isabelle Margaill, Jean-François Berret and Virginie Beray-Berthat*

**Outline**

S1 – X-ray photoelectron spectrometry (XPS) Ce 3d spectra
S2 – Copolymers synthesis and characterization
S3 – Quantification of nanoceria primary amine groups
S4 – Effect of nanoceria on ROS production on bEnd.3 resting cells
S5 – Effect of glutamate and $H_2O_2$ on mitochondrial ROS production

This version: Wednesday, April 7, 2021

**Supplementary Information S1 – X-ray photoelectron spectrometry (XPS) Ce 3d spectra**

XPS data was collected using an Omicron Argus X-ray photoelectron spectrometer using a monochromated $AlK_{\alpha}$ (1486.6 eV) radiation source with a 300 W electron beam power. The emission of photoelectrons from the sample was analyzed at a takeoff angle of 45° under ultra-high vacuum conditions ($10^{-8}$ Pa). The spectra were collected at pass energy of 100 eV for the survey scan and 20 eV for the high-resolution scans.

With XPS , the peaks are due to ejected Ce3d electrons from $Ce^{4+}$ and $Ce^{3+}$ whose states and binding energies are detailed in Table S1. There are 3 peaks associated with $Ce^{4+}$ ions and 2 peaks associated with $Ce^{3+}$ ions each of them split in two, $Ce3d_{5/2}$ ($V^i$) and $Ce3d_{3/2}$ ($U^i$) states, presenting a constant separation of ~ 18.5 eV. The three doublets corresponding to $Ce^{4+}$ are U''' (916.7 eV)/V''' (898.4 eV), U (901.0 eV)/V (882.5 eV), U'' (907.3 eV)/V'' (888.8 eV) and the two doublets corresponding to $Ce^{3+}$ are U' (903.5 eV)/V' (884.9 eV), $U_o$ (898.8 eV)/$V_o$ (880.3 eV)). The spectra were decomposed using the free software XPSPEAKS 4.1. Some conditions were assumed for the deconvolution:

1) Gaussian-Lorentzian curves were used for the individual peaks, the weight of the Gaussian /Lorentzian contribution being optimized for the U''' peak and fixed for all the others.
2) the full width at half maximum (FWHM) of split peaks has the same values. The fraction of $Ce^{3+}$ ions was calculated from the integrated intensities of the XPS peaks through Eq. S2.

The $Ce^{3+}$ percentage is calculated using the expression:

$$f_{Ce^{3+}} = \frac{U' + V' + Uo + Vo}{U''' + V''' + U + V + U'' + V'' + U' + V' + Uo + Vo}$$

Leading to the data in Table S1.

| Ion | State | Binding energy (eV) | CeO$_2$ nanoparticles |
|---|---|---|---|
| Ce$^{4+}$ | U''' | 916.7 | 11933/2.13/13 |
| | V''' | 898.4 | 16579/2.13/13 |
| | U | 901.0 | 11375/1.95/13 |
| | V | 882.5 | 15391/1.95/13 |
| | U'' | 907.3 | 7860/3.83/13 |
| | V'' | 888.8 | 13397/3.83/13 |
| Ce$^{3+}$ | U' | 903.5 | 4213/2.75/13 |
| | V' | 884.9 | 7991/2.75/13 |
| | U$_o$ | 898.8 | 0/-/- |
| | V$_o$ | 880.3 | 0/-/- |
| | Ce$^{3+}$ (%) | | 13.8 |

**Table S1** – Parameters and integrated intensities of peaks from XPS Ce 3d spectra of nanoceria and the Ce$^{3+}$ fraction $f_{Ce^{3+}}$ calculated.

**Supplementary Information S2 – Copolymers synthesis and characterization**

**S2.1 -** MPEG$_{2K}$-MPh

The phosphonic acid PEG copolymers were synthesized by Specific Polymers. Synthesis and characterization of MPEG$_{2K}$-MPh is described in detail in V. Torrisi et al., Biomacromolecules. 15 (2014) 3171–3179.

**S2.2** MPEG$_{2K}$-MPEGa$_{2K}$-MPh

The phosphonic acid PEG terpolymer containing amino functions was synthesized by Specific Polymers. Synthesis and characterization of MPEG$_{2K}$-MPEGa$_{2K}$-MPh is described below.

**S2.2.1. Materials**

Poly(ethylene glycol) mono-functional vinylether (Vinyl-PEG, CAS: 79–41–4) was supplied by Ineos and used as received. Methacryloyl chloride (CAS: 920-46-7, Sigma Aldrich) was distilled before use. Dimethyl(methacryoyloxy)methyl phosphonate (MPh$_e$, SP-41-003, CAS: 86242–61–7) monomers was produced by SPECIFIC POLYMERS. Cysteamine hydrochloride (CAS:156-57-0), Di-tert-butyl dicarbonate (Boc$_2$O, CAS: 24424-99-5), 2,2'-Azobis(isobutyronitrile) (AIBN; CAS: 78–67–1) were supplied by Sigma-Aldrich and used as received.

**S2.2.2. Nomenclature**

A simplified nomenclature has been set up in order to describe the comonomers and polymers synthesized and used in this study. Poly(ethylene glycol) methacrylate monomers were named MPEG. When the MPEG has an amine functional group at the end of the chain it is named MPEGa2K, where the letter 'a' refers to the presence of amino functionality and '2K' to the molecular weight of poly(ethylene glycol) chain (2000 g mol$^{-1}$). When the amine functional group is Boc-protected, it is stated by substituting 'a' by '@'. As for example, MPEG@2K corresponds to a poly(ethylene glycol) methacrylate of 2000 g mol$^{-1}$ bearing amine-Boc end group. MPEG2K corresponds to poly(ethylene glycol) methacrylate of 2000 g mol$^{-1}$ with an inert methoxy end group at the chain ends. In this work, dimethyl(methacryoyloxy)methyl phosphonate monomer was called MPh$_e$. After hydrolysis of the phosphonate ester into phosphonic acid, the corresponding repetitive unit was named MPh. Finally, the polymer named MPEG$_{2K}$-MPEGa$_{2K}$-MPh is a statistic terpolymer of poly(ethylene glycol) methacrylate 2000 g mol$^{-1}$, amino-poly(ethylene glycol) methacrylate 2000 g mol$^{-1}$ and dimethyl(methacryoyloxy) phosphonic acid. Nomenclature used all along this article is summarized in Table 1.

*Table 1. Monomers and polymers simplified nomenclature*

| Name | Simplified naming |
|------|-------------------|
| Poly(ethylene glycol) methacrylate | MPEG |
| Poly(ethylene glycol) amine (ng.mol$^{-1}$) | PEGa2K |
| Poly(ethylene glycol) Boc-amine (ng.mol$^{-1}$) | PEG@2K |
| Poly(ethylene glycol) methacrylate (mg.mol$^{-1}$) | MPEG2K |
| Poly(ethylene glycol) methacrylate Boc-amine (ng.mol$^{-1}$) | MPEG@2K |
| Dimethyl(methacryoyloxy)methyl phosphonate | MPh$_e$ |
| (Methacryoyloxy)methyl phosphonic acid | MPh |

**S2.2.3. Synthesis of precursors and terpolymer**

*Synthesis of the precursor MPEG@2K*

The poly(ethylene glycol) based methacrylate monomer bearing Boc-protected amine group was synthesized within three synthetic steps from commercial Vinyl-PEG. (Fig. S1-a). The amine group is introduced by radical addition of cysteamine on vinyl function of the vinyl-PEG **1** in presence of AIBN. Free amine **2** is obtained by the neutralization of hydrochloride salt under basic condition. Amine is then protected with Boc group to avoid formation of any bifunctional product during the methacryloylation step. The excess of Boc$_2$O is removed with pentane leading to **3**. The monomer MPEG@2K **(4)** is obtained by methacryloylation of hydroxyl group, and precipitate as white powder.

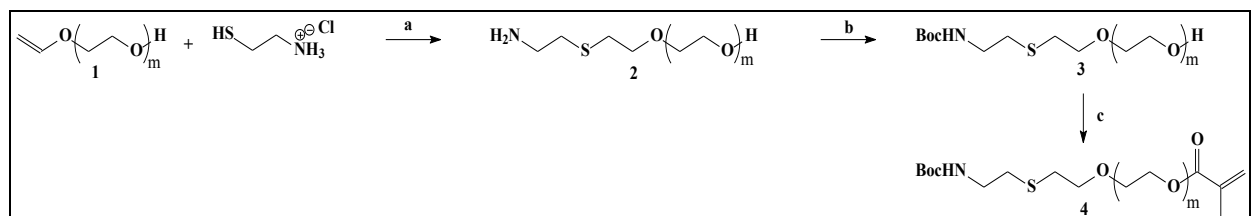

**Figure S2-a**. *Synthesis of MPEG@2K monomer. Reagents and conditions: (a) AIBN, Ethanol, 70°C, 78%; (b) Boc$_2$O, tetrahydrofuran, rt, 95%; (c) Methacryloyl chloride, Net$_3$, CH$_2$Cl$_2$, 90%.*

*PEGa2K **(2)***. Vinyl-PEG (10 g, 5 mmol, 0.2 eq) and cysteamine hydrochloride (59.9 g, 0.53 mol, 20 eq) was introduced in a two-necked round-bottom flask containing 150 mL of ethanol. The reaction mixture was stirred with a magnetic stirrer and slightly heated till solubilization. The reaction mixture was degassed with argon. Half of the AIBN (2.1 g, 13 mmol, 0.5 eq) was added to the mixture and the overall reactive media was heated to 70 °C. In the meantime, a solution of vinyl-PEG (42.7 g, 21 mmol, 0.8 eq) in ethanol (50 ml) was prepared with the rest of AIBN (2.1 g, 13 mmol, 0.5 eq) and degassed under argon. Vinyl-PEG solution was added dropwise for an hour. The reactive media was maintained at 70°C for 24 h leading to 100% conversion. The vinyl-PEG conversion was measured by $^1$H NMR spectroscopy. Ethanol was removed by evaporation under reduce pressure. The reactive media was dissolved into dichloromethane and aqueous and basic extractions were performed. The mixture was dried with Na$_2$SO$_4$, concentrated under vacuum then solubilized in a minimum of THF to be precipitated in cold ether. PEGa2K, **2** was obtained as a white powder. Yield 78%. $^1$H NMR (300 MHz, D$_2$O) δ (ppm): 3.72 − 3.38 (C$\underline{H}_2$-C$\underline{H}_2$-O), 2.71 − 2.60 (C$\underline{H}_2$-S-C$\underline{H}_2$), 2.58 − 2.50 (C$\underline{H}_2$-NH$_2$). Its $^1$H NMR spectrum is shown in Fig. S1-c.

*PEG@2K **(3)***. A solution of di-tert-butyl dicarbonate (Boc$_2$O, 4.8g, 22mmol, 1.1eq) in THF (30 mL) was slowly added to a solution of PEGa2K (**2**, 43g, 20.2mmol) in tetrahydrofuran (THF, 300 mL) under stirring. The mixture was stirred for 12 h and then concentrated under reduce pressure. The solid obtained was washed with pentane to remove the excess of Boc$_2$O, filtrated and dried to obtain PEG@2K (**3**) as a white powder. Yield 95%. $^1$H NMR (300 MHz, D$_2$O) δ (ppm): 3.72 − 3.36 (C$\underline{H}_2$-C$\underline{H}_2$-O), 3.13 (C$\underline{H}_2$-S-CH$_2$), 2.64 (CH$_2$-S-C$\underline{H}_2$), 2.55 (C$\underline{H}_2$-NH$_2$), 1.31 (C(C$\underline{H}_3$)$_3$). Its RMN spectrum is shown in Fig. S1-d.

*MPEG@2K **(4)***. Methacryloyl chloride (2.2g, 21mmol, 1.1eq) was added dropwise to a solution of PEG@2K (**3**, 42.4g, 19mmol), triethylamine (2.3g, 23mmol, 1.2eq) and dichloromethane (50 mL) at 0°C under stirring. The solution was stirred at room temperature for 4 hours. The reactive media was recovered and washed under acidic and basic conditions. The product is washed with water then dried with Na$_2$SO$_4$ and concentrated under reduce pressure. The wax is solubilized in a minimum of THF then precipitated in cold ether. MPEG@2K (**4**) is obtained as a white powder. Yield 90%. $^1$H NMR (300 MHz, D$_2$O) δ (ppm): 6.04 (O=C-C(C$\underline{H_2}$)-CH$_3$), 5.62 (O=C-C(C$\underline{H}$$_3$)-CH$_3$), 4.30 − 4.17 (CH$_2$-C$\underline{H_2}$-O-C=O), 3.76 − 3.39 (C$\underline{H_2}$-C$\underline{H_2}$-O), 3.16 (C$\underline{H_2}$-NH$_2$), 2.67 (C$\underline{H_2}$-S-CH$_2$), 2.58 (CH$_2$-S-C$\underline{H_2}$), 1.82 (O=C-C(CH$_2$)-C$\underline{H_3}$), 1.34 (C(C$\underline{H_3}$)$_3$). Its $^1$H NMR spectrum is shown in Fig. S1-e.

*Synthesis of MPEG$_{2K}$–MPEGa$_{2K}$-MPh terpolymer*

Free radical polymerization of MPEG@2K (**4**), MPEG2K and MPh$_e$ with AIBN as radical initiator was performed leading to the targeted terpolymers (Fig. S1-b). The deprotection of functional Boc-protected amino groups and phosphonated esters of terpolymer MPEG2K-MPEG@2K-MPh$_e$ (**5**) were carried out simultaneously in presence of bromotrimethylsilane (BrSiMe$_3$). (A. Graillot et al., Polym. Chem. 2013, 4, 795−803. B. Canniccioni et al., Polym. Chem. 2013, 4, 3676−3685.) Following this step, phosphonic acid and amine groups were recovered. MPEG$_{2K}$-MPEGa$_{2K}$-MPh (**6**) was obtained by precipitation in cold ether as a slightly brown powder. A representative synthesis targeting MPEG$_{2K}$-MPEGa$_{2K}$-MPh is described.

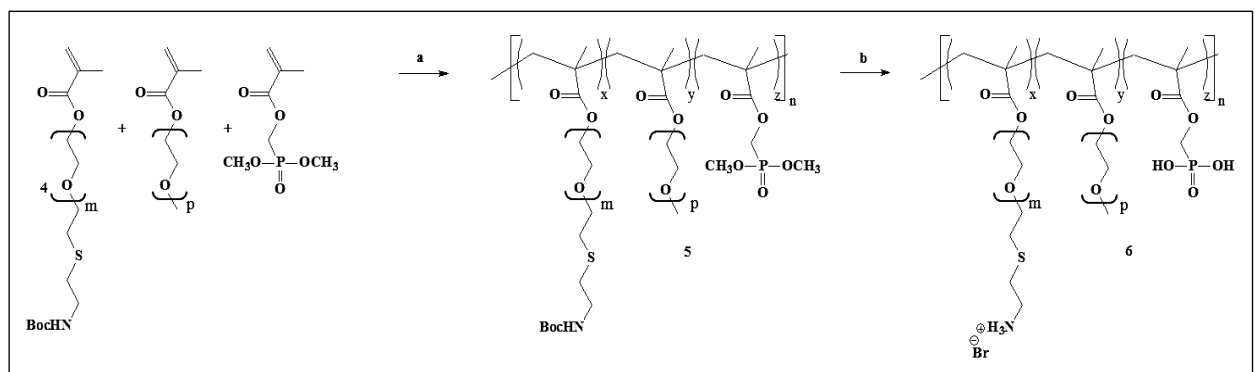

**Figure S2-b**. *Synthesis of MPEG$_{2K}$–MPEGa$_{2K}$-MPh terpolymer. Reagents and conditions: (a) AIBN, methyl ethyl ketone, 70°C, 85%; (b) BrSiMe$_3$, CH$_2$Cl$_2$, EtOH, 84%.*

*MPEG2K-MPEG@2K-MPh$_e$ (**5**).* MPh$_e$ (1.14g, 5.5 mmol), MPEG2K (1.5 g, 0.8 mmol), MPEG@2K (10.3g, 4.7 mmol) and AIBN (0.03 g, 0.16 mmol) were added along with 50 mL of methyl ethyl ketone (MEK) in a two-necked round-bottom flask. The mixture was heated at 70 °C under argon in a thermostatic oil bath for 24 hours leading to 100% conversion. Conversion was monitored by $^1$H NMR spectroscopy. MEK was evaporated and the terpolymer was dissolved in a small volume of THF before precipitation in cold ether. MPEG2K-MPEG@2K-MPh$_e$ (**5**) was obtained as a white powder (85% yield). $^1$H NMR (300 MHz, D$_2$O) δ (ppm): 4.49 − 3.91 (C$\underline{H_2}$-O-C=O), 3.90 − 3.27 (C$\underline{H_2}$-C$\underline{H_2}$-O, O=P-(OCH$_3$)$_2$), 3.22 (CH$_2$-O-C$\underline{H_3}$), 3.13 (C$\underline{H_2}$-NH$_2$), 2.63 (C$\underline{H_2}$-S-CH$_2$), 2.55 (CH$_2$-S-C$\underline{H_2}$), 2.19 − 0.60 (C(C$\underline{H_3}$)-C$\underline{H_2}$), 1.32(C(C$\underline{H_3}$)$_3$).$^{31}$P NMR (D$_2$O, 300 MHz) δ (ppm): 23.4 (O=$\underline{P}$-(OCH$_3$)$_2$). Its $^1$H NMR spectrum is shown in Fig. S1-f.

MPEG$_{2K}$-MPEGa$_{2K}$-MPh (**6**). Previously obtained terpolymer (MPEG2K-MPEG@2K-MPh$_e$, **5**, 11g) was dissolved in a minimum of dichloromethane (10 mL) and degassed with argon. Bromotrimethylsilane (6.7g, 44 mmol) was added dropwise to the reactive media under inert and dry conditions. The solution was stirred for 8 h, then the solvent was removed with a rotatory evaporator. Ethanol was then added in excess to complete the ethanolysis. Finally, MPEG$_{2K}$-MPEGa$_{2K}$-MPh was obtained by precipitation in cold ether (84% yield). $^1$H NMR (300 MHz, D$_2$O) δ (ppm): 4.33 − 3.27 (C$\underline{H_2}$-O-C=O, CH$_2$-C$\underline{H_2}$-O), 3.23 (CH$_2$-O-C$\underline{H_3}$), 3.08 (C$\underline{H_2}$-NH$_2$), 2.74 (C$\underline{H_2}$-S-

CH$_2$), 2.66 (CH$_2$-S-C$\underline{H}_2$), 2.26 − 0.46 (C(C$\underline{H}_3$)-C$\underline{H}_2$).$^{31}$P NMR (D$_2$O, 300 MHz) δ (ppm): 15.25 (O=$\underline{P}$-(OH)$_2$). Its $^1$H NMR spectrum is shown in Fig. S1-g

## S2.2.4. RMN spectra

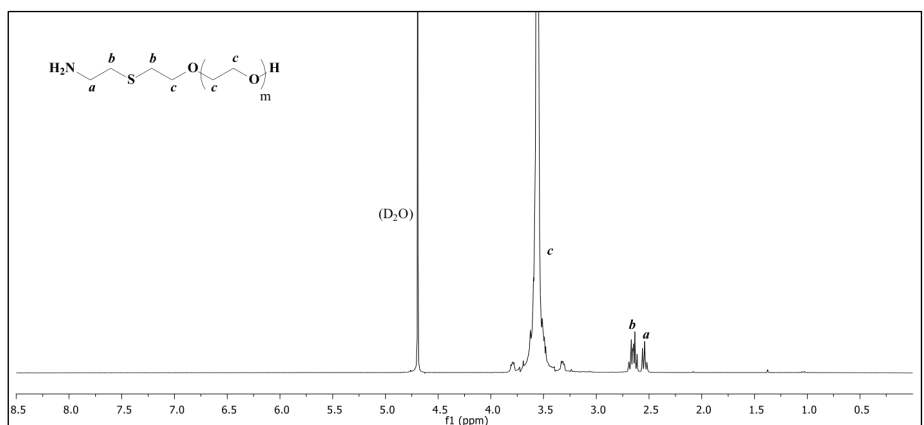

**Fig. S2-c.** *$^1$H NMR spectrum of PEGα2K. $^1$H NMR (300 MHz, D$_2$O) δ (ppm): 3.72 − 3.38 (C$\underline{H}_2$-C$\underline{H}_2$-O), 2.71 − 2.60 (C$\underline{H}_2$-S-C$\underline{H}_2$), 2.58 − 2.50 (C$\underline{H}_2$-NH$_2$).*

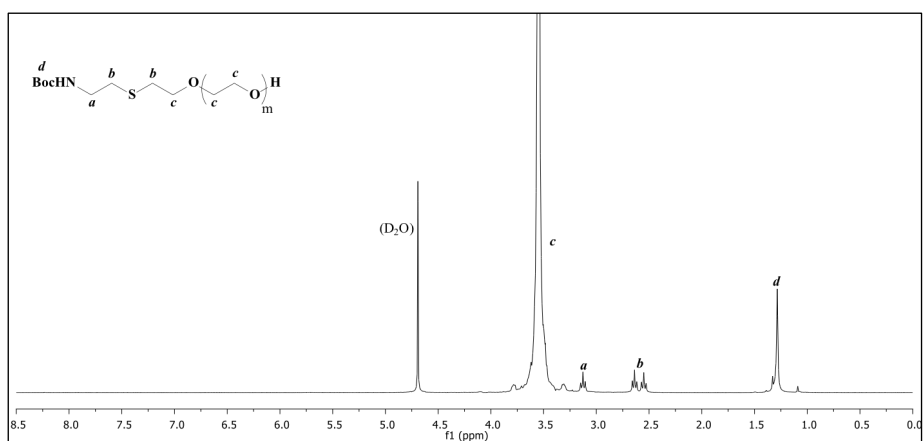

**Fig. S2-d.** *$^1$H NMR spectrum of PEG@2K. $^1$H NMR (300 MHz, D$_2$O) δ (ppm): 3.72 − 3.36 (C$\underline{H}_2$-C$\underline{H}_2$-O), 3.13 (C$\underline{H}_2$-S-CH$_2$), 2.64 (CH$_2$-S-C$\underline{H}_2$), 2.55 (C$\underline{H}_2$-NH$_2$), 1.31 (C(C$\underline{H}_3$)$_3$).*

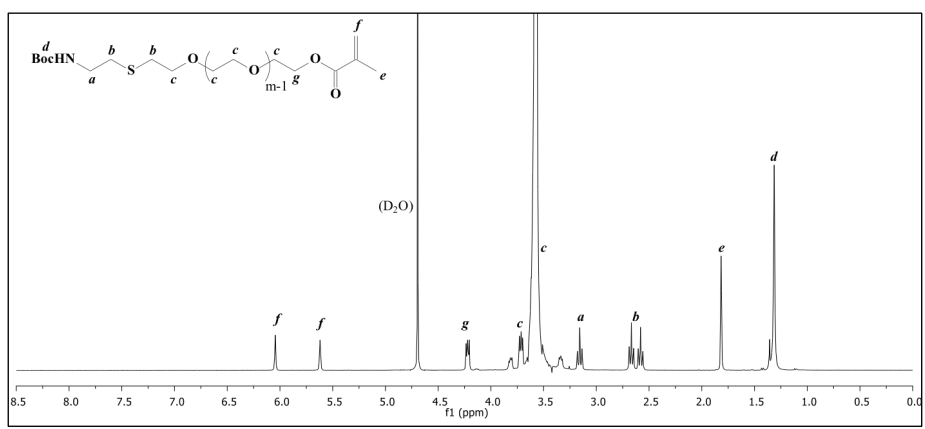

**Fig. S2-e.** *¹H NMR spectrum of MPEG@2K. ¹H NMR (300 MHz, D₂O) δ (ppm): 6.04 (O=C-C(C$\underline{H}_2$)-CH₃), 5.62 (O=C-C(C$\underline{H}_2$)-CH₃), 4.30 − 4.17 (CH₂-C$\underline{H}_2$-O-C=O), 3.76 − 3.39 (C$\underline{H}_2$-C$\underline{H}_2$-O), 3.16 (C$\underline{H}_2$-NH₂), 2.67 (C$\underline{H}_2$-S-CH₂), 2.58 (CH₂-S-C$\underline{H}_2$), 1.82 (O=C-C(CH₂)-C$\underline{H}_3$), 1.34 (C(C$\underline{H}_3$)₃).*

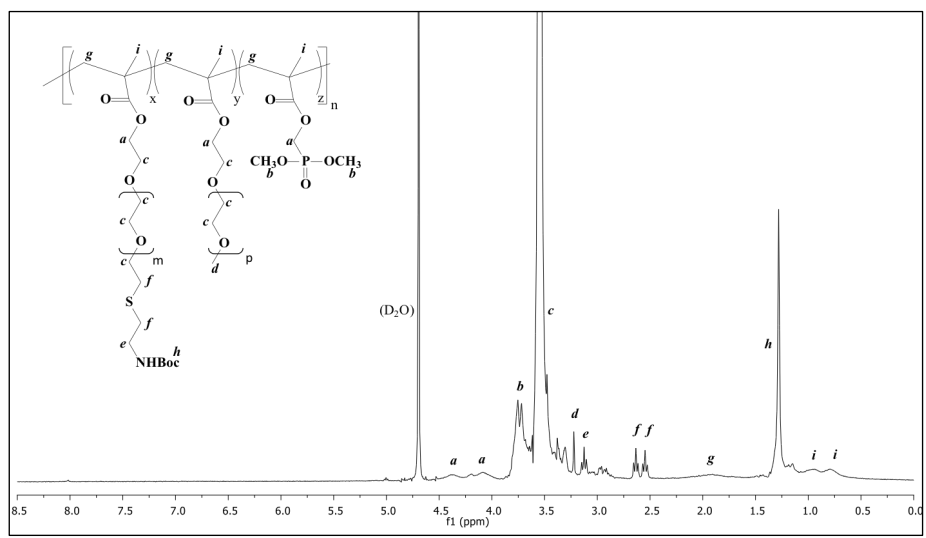

**Fig. S2-f.** *¹H NMR spectrum of MPEG2K-MPEG@2K-MPh$_e$. ¹H NMR (300 MHz, D₂O) δ (ppm): 4.49 − 3.91 (C$\underline{H}_2$-O-C=O), 3.90 − 3.27 (C$\underline{H}_2$-C$\underline{H}_2$-O, O=P-(OCH₃)₂), 3.22 (CH₂-O-C$\underline{H}_3$), 3.13 (C$\underline{H}_2$-NH₂), 2.63 (C$\underline{H}_2$-S-CH₂), 2.55 (CH₂-S-C$\underline{H}_2$), 2.19 − 0.60 (C(C$\underline{H}_3$)-C$\underline{H}_2$), 1.32(C(C$\underline{H}_3$)₃).*

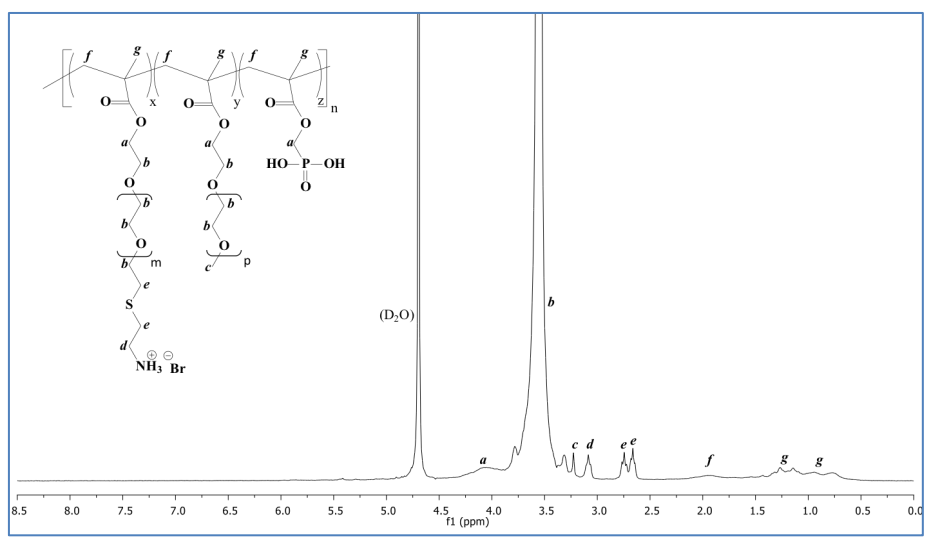

**Fig. S2-g.** *¹H NMR spectrum of MPEG$_{2K}$-MPEGα$_{2K}$-MPh. ¹H NMR (300 MHz, D₂O) δ (ppm): 4.33 − 3.27 (C$\underline{H}_2$-O-C=O, C$\underline{H}_2$-C$\underline{H}_2$-O), 3.23 (CH₂-O-C$\underline{H}_3$), 3.08 (C$\underline{H}_2$-NH₂), 2.74 (C$\underline{H}_2$-S-CH₂), 2.66 (CH₂-S-C$\underline{H}_2$), 2.26 − 0.46 (C(C$\underline{H}_3$)-C$\underline{H}_2$).*

**Supplementary Information S3 – Stability of polymer coated cerium oxide nanoparticles in cell culture medium**

To assess the coating performances of the PEGylated polymers in physiological solvents, light scattering was performed to test the dispersion state over time. In the experiments performed, 20 µL of a 20 g L-1 nanoparticle dispersion are diluted ten times in complete cell culture medium. The dispersions are then studied as a function of time at day 1, 2 and 7. A final assessment is realized at 60 days after mixing.

Figure S3 display the second-order autocorrelation function of the scattered light, $g^{(2)}(t)$ obtained in cell culture medium for $CeO_2$ coated with the phosphonic acid PEG polymers (left column) and copolymers (right column). The PEG molecular weights are 2 KDa in the first line, 5 KDa in the second line, whereas the third line shows results for the amine terminated PEG copolymers.

At day 2 and 7 (Fig. S3a and c), $CeO_2$@PEG2K-Ph and $CeO_2$@PEG5K-Ph show signs of aggregation, with autocorrelation function $g^{(2)}(t)$ shifting to the right hand-side and corresponding to the decrease of the diffusion constant. The intensity distributions in the insets also illustrate this augmentation. For the PEG5K-Ph coat, the aggregation is seen only after two months. For the monophosphonic acid PEG polymers, the displacement of the PEGnK-Ph away from the surface probably favors the protein adsorption, which again accelerates the destabilization kinetics.

In contrast, nanoceria coated with the multi-functionalized copolymers have $g^{(2)}(t)$ that remain unchanged over time (Figure SI3b, d and e). The autocorrelation functions exhibit again a unique relaxation mode at day 1, 2 and 7 associated hydrodynamic diameters of 27.2, 31.7 and 31.5 nm. For these samples, the brush thickness is not altered by the presence of proteins and remains at the value found in DI-water, 8.9, 11.8 and 11.3 nm, respectively. These outcomes suggest that the coated nanoceria are devoid of a protein corona.

Data taken at 60 days are provided in Table SI3, corroborating the above results.

**Table S3**. *Time evolution of the hydrodynamic diameter $D_H$ from polymer coated nanoceria dispersed in complete DMEM. Particle aggregation is denoted in red.*

| Particles | Hydrodynamic Diameter $D_H$ (nm) | | | |
|---|---|---|---|---|
| | Day 1 | Day 2 | Day 7 | Day 60 |
| $CeO_2$@PEG$_{2K}$-Ph | 12 | 17 | 51 | >1000 |
| $CeO_2$@PEG$_{5K}$-Ph | 11 | 12 | 24 | 207 |
| $CeO_2$@MPEG$_{2K}$-MPEGa$_{2K}$-MPh | 30 | 33 | 26 | 20 |

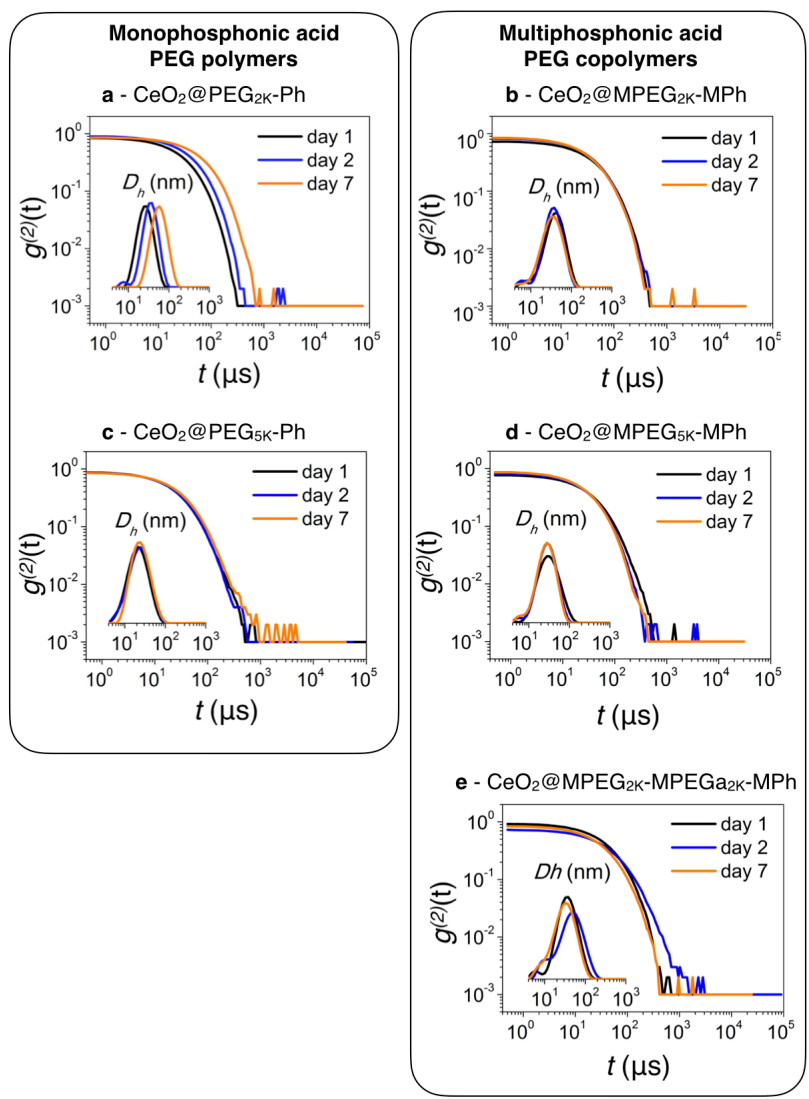

***Figure S3:*** *Autocorrelation functions $g^{(2)}(t)$ obtained from dynamic light scattering on phosphonic acid PEG polymer coated CeO$_2$ nanoparticles as a function of the time. Experiments were performed in cell culture medium (Dulbecco's Modified Eagle's Medium, DMEM) complemented with 10% fetal bovine serum (FBS): **a)** PEG$_{2K}$-Ph; **b)** PEG$_{5K}$-Ph; **c)** MPEG$_{2K}$-MPh; **d)** MPEG$_{5K}$-MPh; **e)** MPEG$_{2K}$-MPEGa$_{2K}$-MPh. The insets display the intensity distributions for the hydrodynamic diameter $D_H$.*

**Supplementary Information S4 – Quantification of nanoceria primary amine groups**

The primary amine groups from CeO$_2$@P2 and CeO$_2$@P3 were quantified by spectrofluorometry using fluorescamine (43749.ME, VWR, Fontenay-sous-Bois, France).

The primary amine groups from CeO$_2$@P2 and CeO$_2$@P3 were quantified by spectrofluorometry using fluorescamine (43749.ME, VWR, Fontenay-sous-Bois, France). Glycine (3908.2, Roth, Lauterbourg, France) which has a single primary amine function was used to achieve a standard of 1 μM to 1 mM. The standard and samples are duplicated at 100μL per well in a 96-well plate. A solution of fluorescamine (43749.ME, VWR, Fontenay-sous-Bois, France) is prepared at 3 mg.ml$^{-1}$ in acetone (9372.1, Roth), 50 μl of reagent are added per well. The reading is carried out with the fluorimeter (Infinite F200 PRO, TECAN, Lyon, France) at an excitation wavelength of 400 nm and emission of 460 nm.

### Standard curve of glycine

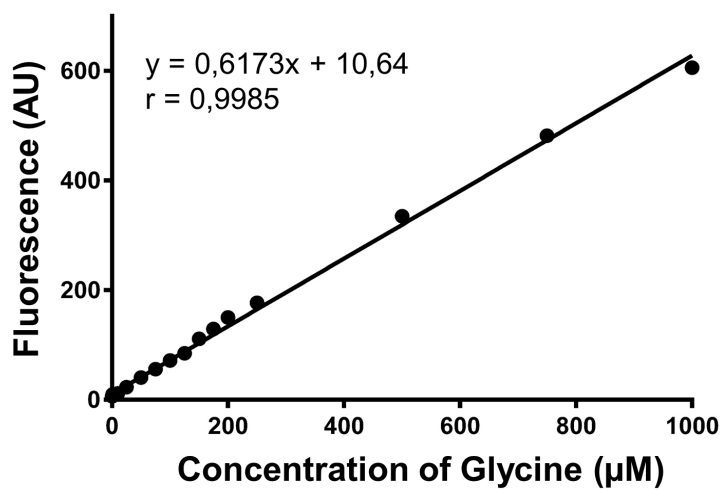

y = 0,6173x + 10,64
r = 0,9985

| | CeO$_2$@P2 | CeO$_2$@P3 |
|---|---|---|
| Fluorescence (AU) | 29,5 | 80,0 |
| Primary amine (µM) | 30,6 | 112,4 |

**Figure S4: Determination primary amine groups from CeO2@P2 and CeO2@P3.**

**Supplementary Information S5 – Effect of nanoceria on ROS production on bEnd.3 resting cells**

The effect of nanoceria on ROS production was assessed on bEnd.3 resting cells. Antioxidant capacities were measured using the dichlorofluorescein diacetate ($H_2DCFDA$) probe after 4 and 24 hours of incubation (Figure S5). The antioxidant NAC significantly reduced the basal levels of intracellular ROS by at least 50% at both incubation times ($P<0.01$, $P<0.001$). After a 4-hour incubation, CNPs did not decrease basal ROS production of bEnd.3 resting cells except for $CeO_2@P1$ at 100 $\mu g.mL^{-1}$ (-22%, $P<0.05$) and $CeO_2@P3$ at 1000 $\mu g.mL^{-1}$ (-21%, $P<0.01$). After 24 hours of incubation, none of the CNPs reduced ROS production.

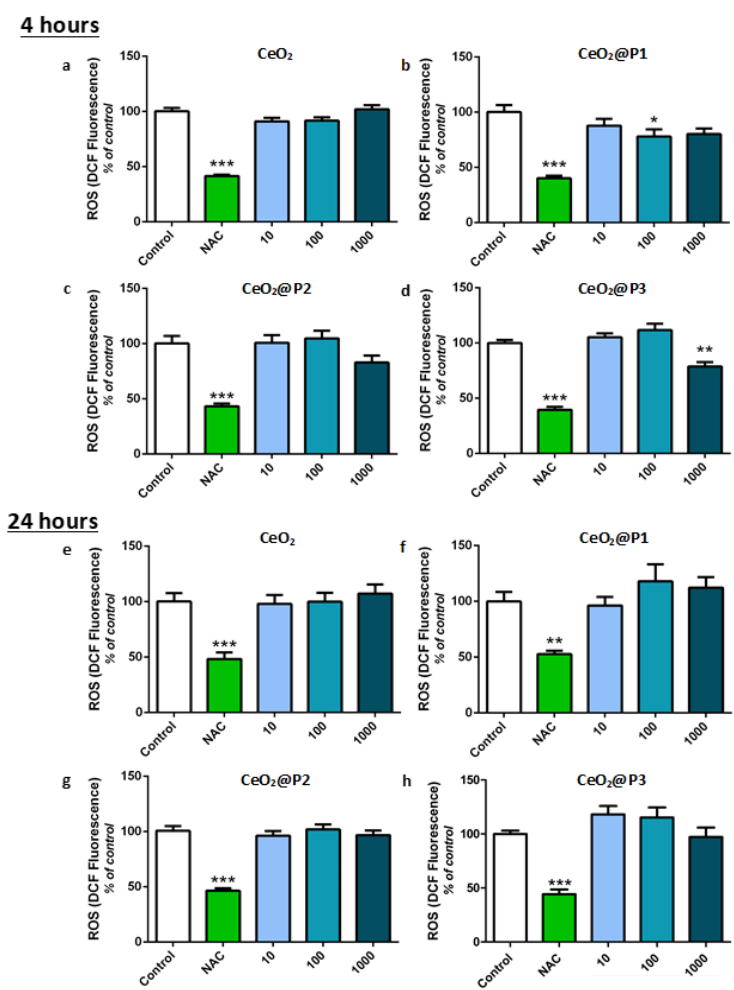

**Figure S5:** Species wer ... ive Oxygen and 24 hou ... hours (a-d) control cell ... rescence of nanoceria (a ... mM. Bare 1000 $\mu g.mL$ ... 10, 100 and ... rs, * $P<0.05$, ** $P<0.01$, *

**Supplementary Information S6 – Effect of glutamate and $H_2O_2$ on mitochondrial ROS production**

The mitochondrial production of superoxide anions was assessed using the MitoSOX™ Red reagent. bEnd.3 cells were seeded in 24-well plates at a density of 200,000 cells per well in DMEM medium and cultured for 24 hours. The cells were treated for 4 hours with DMEM (control cells), glutamate (100 mM) or $H_2O_2$ (2 mM).

In control condition, the percentage of MitoSOX™ Red reagent-positive cells was 1,1 ± 0,1% and was not modified by glutamate treatment (**Figure S5**). By contrast, $H_2O_2$ significantly increased the number of labelled cells (P < 0.01).

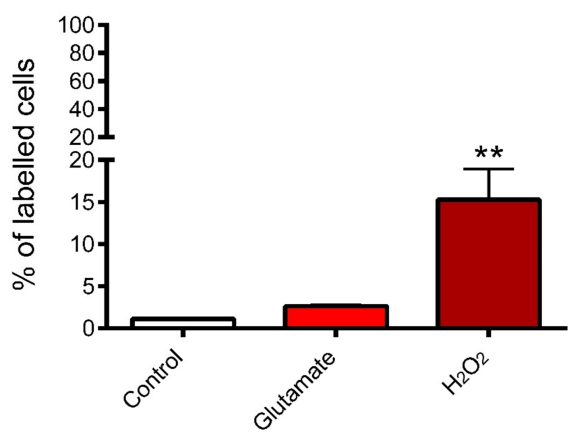

**Figure S6: Effect of glutamate and $H_2O_2$ on mitochondrial ROS production.** bEnd.3 cells were treated for 4 hours with DMEM (control cells), glutamate (100 mM) or $H_2O_2$ (2 mM). Results are expressed as a percentage of positive cells. Data expressed as mean ± SEM, n=3. ANOVA and Dunett's test, ** P < 0.01 *versus* control.